\documentstyle[aps,floats,eqsecnum,epsfig]{revtex}
\begin{document}
\draft
\def\D{{\cal D}}
\def\S{\cal S}
\def\f{\vert f\vert}
\def\tr{{\rm Tr}}
\def\al{$\alpha$}
\def\de{\partial}
\def\fhi{\varphi}
\def\s{$s^{\mu}$}
\def\half{1\over 2}
\def\c{\cos}
\def\s{\sin}
\def\be{\begin{equation}}
\def\ee{\end{equation}}
\def\bea{\begin{eqnarray}}
\def\eea{\end{eqnarray}}
\def\i{$\infty$}
\def\mi{$-\infty$}
\def\ip{$\int_{\-\infty}^{\infty}$}
\def\r{\hat{r}}
\def\m{\hat{m}}
\def\n{\hat{n}}
\def\x{\hat x}
\def\y{\hat y}
\def\z{\hat r}
\def\O{\hat\Omega}
\def\k{\hat k}
\def\X{\hat X}
\def\Y{\hat Y}
\def\Z{\hat Z}
\def\d{\rm d}
\newcommand{\bd}{\begin{displaymath}}
\newcommand{\ed}{\end{displaymath}}
\newcommand{\prim}{{\scriptscriptstyle \prime}}
\newcommand{\bv}[1]{\mathbf[#1]}

\title{On the Detection of a Scalar Stochastic Background of Gravitational
Waves}

\author{Danilo Babusci}
\address{INFN- Laboratori Nazionali di Frascati, I-00044 Frascati,
Italy}
\author{Luca Baiotti}
\address{SISSA and INFN sez. di Trieste,\\
Via Beirut 4, 34013 Trieste, Italy}
\author{Francesco Fucito}
\address{INFN sez. di Roma 2,\\
Via della Ricerca Scientifica, 00133 Roma, Italy}
\author{Alessandro Nagar}
\address{Dipartimento di Fisica, Universit\`a di Parma \\
and INFN, Gruppo Collegato di Parma, 43100 Parma, Italy}
\maketitle

\begin{abstract}
In the near future we will witness the coming to a full operational
regime of laser interferometers and resonant mass detectors of spherical shape.
In this work we study the sensitivity of pairs of such gravitational 
wave detectors to a scalar stochastic background of gravitational waves. 
Our computations are carried
out both for minimal and non minimal coupling of the scalar fields.

\end{abstract}

\section{Introduction}
In few years, the research on the detection of gravitational waves (GWs) 
will hopefully greatly progress. This hope is based on the coming into 
operation
of a new generation of experimental devices that, if they can be operated 
at the 
planned sensitivity, should probe deeply into the region in which we believe
GWs can be observed. According to the experimental technique employed, these 
detectors can be divided into two categories: interferometric detectors and 
resonant mass detectors. 
To make our point more concrete let us concentrate on Michelson 
interferometers
\cite{ligo,virgo} and resonant mass detectors of spherical shape \cite{odia}.
The main advantage of interferometers is their sensitivity in a wide frequency
band. On the other hand spherical shaped resonant mass detectors at resonance
have the same sensitivity regardless of the direction of the impinging GW.

In the following we will concentrate on a very specific issue, that is on
the possibility of detecting scalar GWs. Our interest in this subject stems
from the observation that Einstein's gravity is definitively not the only 
mathematically consistent theory
of gravity and in fact the presence of scalar fields coupled to gravity is 
required by a vast array of theories that model various phenomena
as the inflationary universe or attempt to incorporate gravity with
the quantum world. For a review on this subject see Ref. \cite{will}.
For more recent proposals that also require a modification of Einstein's 
gravity see Refs. \cite{add,aadd,add1}. Are all the above described detectors 
fit to measure scalar GWs? While the answer is obvious for resonant mass detector
of spherical shape, the situation for interferometers must be analyzed with care.
Let us use for a moment the ``standard'' description, that is well suited for 
our kind of
argument, of an impinging GW (for the moment we neglect its spin content and 
direction)  stretching the lengths $L_1$ and $L_2$ of the two arms of 
the interferometer. 
The conventional Michelson interferometer is configured for maximizing 
its sensitivity in the detection of the {\it differential mode} signal 
$\Delta_-=\delta L_1 - \delta L_2$. Even if the information regarding
$L_1$ and $L_2$ separately is available, the sensitivity of these measurements
is orders of magnitude worse than that of $\Delta_-$, and, thus, a single 
interferometer of this type is not able to disentangle the {\it common mode} 
signal $\Delta_+=\delta L_1+\delta L_2$ (transverse monopole mode) from 
$\Delta_-$ (usual spin 2 mode). A way out could be 
the construction of an array of these detectors or the adoption of 
a different optical configuration (Fox-Smith) for the 
interferometer \cite{tobar}. Even if interesting from a theoretical point of 
view, these alternatives
do not seem practical, given the cost and the difficulty in operating such 
complex
apparatus. A viable alternative to these proposals could be, from our point of view, 
that of a coincidence analysis on the data of an interferometer and a resonant
mass detector of spherical shape \cite{maggiore}.

In this work we study the sensitivity of combined pairs of resonant mass detectors and 
interferometers to a scalar stochastic background of gravitational waves (SBGW).
If such a background has a flat spectrum (which is the
standard assumption) even the narrow frequency band available to a resonant mass 
detector won't have much influence on our conclusions. Our computations generalize
the results of Ref. \cite{maggiore} in which the sensitivity patterns
to scalar radiation were considered. To be as general as possible, the impinging
radiation is computed in the  general setting given by  
scalar tensor theories \cite{Damour}. Our main result is the computation of the sensitivity
to scalar GWs of correlated pairs of (solid mass or hollow)
resonant  mass detectors of spherical shape or pairs of 
interferometer-resonant mass detector.
Finally we consider the effects on such detectors of 
massless non minimally coupled scalar fields, generalizing the results 
of Refs. \cite{g1,g2}. While this paper was being written, a similar analysis 
employing two LIGO interferometers for massive and nonrelativistic scalar particles
 appeared \cite{g3}.

\section{Scalar tensor theory}
\subsection{Fundamental equations}
Let us consider a very general tensor multi scalar theory of gravity, where
the gravitational interaction is mediated by $n$ long range scalar fields 
$\varphi^{a}$ 
in addition to the usual tensor field present in Einstein's theory. The action in the 
Einstein frame is
\be
S=\left(16\pi G\right)^{-1}\int {\d}^4 x \sqrt{-g}\,\bigl(R-2g^{\mu\nu}\gamma_{ab}(\varphi^c)
\de_{\mu}\varphi^a\de_{\nu}\varphi^b\bigr)+S_{m}[\Psi_{m},A^2(\varphi^a)g_{\mu\nu}]\;.
\ee
We use units in which the speed of light is $c=1$ and the signature is $- + + + $. 
Greek indices $\lambda,\mu,\nu,\dots=0,1,2,3$ denote spacetime indices; Latin indices 
from the second part of the alphabet $i,j,k,l\dots=1,2,3$ denote spatial indices; 
Latin indices from the first part of the alphabet $a,b,c,\dots=1,\dots, n$   
label the $n$ scalar fields. Our curvature conventions follow those of Ref. 
\cite{Weinberg}. $R=g^{\mu\nu}R_{\mu\nu}$ is the curvature scalar of the Einstein 
metric $g_{\mu\nu}$ and $g=\mathrm{det}(g_{\mu\nu})$. The action contains a 
dimensionful constant $G$, which will be denoted as the bare gravitational 
constant (related to $\tilde{G}$ Newton's constant as measured by Cavendish experiments) 
and a $\sigma$ model type metric $\gamma_{ab}(\varphi)$, not necessarily positive definite, 
in the $n$ dimensional space of the scalar fields. $S_{m}$ denotes the matter action, 
which is a functional of some matter variables $\Psi_{m}$, and of the Jordan-Fierz 
metric $\tilde{g}_{\mu\nu} \equiv A^{2}(\varphi)g_{\mu\nu}$. The scalar fields 
can be non minimally coupled to matter. This means that they can appear 
as coupling ``constants"
between the matter fields $\Psi_{m}$ and gravity $\tilde{g}_{\mu\nu}$. For instance, 
low energy string type theories naturally introduce in the action terms with
couplings of the kind
\be
S_{dil}=-\frac{\beta}{4}\int {\d}^4 x\sqrt{-\tilde{g}}\,\varphi\,
F^{A}_{\mu\nu}F^{A}_{\alpha\beta}
\tilde{g}^{\mu\nu}\tilde{g}^{\alpha\beta},
\ee
where $F^{A}_{\mu\nu}=\de_{\mu}A^{A}_{\nu}-\de_{\nu}{A}_{\mu}^A+
f^{ABC}A^{B}_{\mu}A^{C}_{\nu}$ is the 
Yang-Mills field strength and the scalar field $\varphi$ is the dilaton.

By varying the action $S$ with respect to the Einstein metric $g_{\mu\nu}$ and the scalar 
fields $\varphi^{a}$, one obtains the following field equations
\bea
\nonumber{}\\
\label{fieldtensor}R_{\mu\nu}-{1\over 2}\,R g_{\mu\nu}\,=\,
2\gamma_{ab}(\varphi)\Bigl(\de_{\mu}\varphi^{a}\de_{\nu}\varphi^{b}-
{1\over 2}g_{\mu\nu}g^{\rho\sigma}\de_{\rho}\varphi^{a}\de_{\sigma}\varphi^{b}\Bigr)+
8\pi GT_{\mu\nu},\\
\label{fieldscalar}g^{\mu\nu}\nabla_{\mu}\nabla_{\nu}\varphi^{a}+g^{\mu\nu}
\gamma_{bc}^{a}(\varphi)
\de_{\mu}\varphi^{b}\de_{\nu}\varphi^{c}=-4\pi G\Bigl[\alpha^{a}(\varphi)T+\sigma^{a}\Bigr],
\eea
where $\gamma^{a}_{bc}$ are 
the Christoffel symbols of the metric $\gamma_{ab}(\varphi)$. The functions $\alpha_{a}
(\varphi)\equiv \de_{a}\ln A(\varphi)$ represent the field dependent couplings 
between scalar fields and matter within the metric sector of the theory. 
$T^{\mu\nu}= 2\left(-g\right)^{-1/2}\delta S_{m}/\delta g_{\mu\nu}$ is the stress energy 
tensor, $T$ its trace and $\sigma_{a}=\left(-g\right)^{-1/2}\delta S_{m}/\delta \varphi^{a}$
is the density of scalar charge.
In the Jordan-Fierz frame we would have
\be
\label{passaggio}T_{\mu\nu}=A^{2}(\varphi)\tilde{T}_{\mu\nu},
\qquad\qquad\sigma_{a}=A^{4}
(\varphi)\tilde{\sigma}_{a}\;,
\ee
as can easily be found from their definition \cite{Damour}. Actually, since 
\hbox{$\sqrt{-\tilde{g}}=A^{4}(\varphi)\sqrt{-g}$}, and
\be
\delta_{\varphi}S_{m}=\int {\d}^4 x \,\sqrt{-g}\,\sigma_{a}\,\delta\varphi^{a}=
\int {\d}^4 x\,\sqrt{-\tilde{g}}\,\tilde{\sigma}_{a}\,\delta\varphi^{a}\;,
\ee
one recovers immediately (\ref{passaggio}).

In the literature, scalar tensor theories with $\sigma_{a}=0$ 
(i.e. metric theories) have been studied by many authors, from 
the pioneering work of Jordan, Fierz, Brans, Dicke, Wagoner \cite{Brans} 
to the recent studies of Damour and Esposito-Far\`ese \cite{Damour}. 
This interest arises from the fact that they do not violate the Weak 
Equivalence Principle and so imply geodesic dynamics for neutral
weakly self gravitating bodies. However this is not the most general 
framework, in particular it is not the case of the interesting scalar 
fields foreseen by string theory. For a recent analysis see Ref. \cite{g1}.

Let us compute the expression of the relative acceleration between two weakly 
self gravitating bodies in the general $n$ scalar theory; this formula will 
be the starting point to write the response of a GW detector to a scalar 
tensor wave.

When $\sigma_{a}\neq 0$, the stress energy conservation law in Einstein units 
is \cite{Damour}
\be\label{heart}
\nabla_{\nu}T^{\mu\nu}=\alpha_{a}\nabla^{\mu}\varphi^{a}T-
\sigma_{a}\nabla^{\mu}\varphi^{a}\;,
\ee
or, in the Jordan-Fierz frame
\be
\label{conservation}\tilde{\nabla}^{\nu}\tilde{T}_{\mu\nu}+\tilde{\sigma}_a
\tilde{\nabla}_{\mu}\varphi^a=0.
\ee
This equation implies a non geodesic motion of test mass bodies. This result 
corresponds, for a single scalar field and a particular choice of the coupling 
function $A(\varphi)$, to the lowest order gravidilaton effective action of 
string theory \cite{g1}. However, if $\tilde{\sigma}_{a}=0$ we have 
$\tilde{\nabla}^{\nu}\tilde{T}_{\mu\nu}=\,0$ and so geodesic motion of test 
mass bodies is recovered.

In Ref. \cite{g1}, starting from the single field string like case 
of (\ref{conservation}) the equation of motion of test mass bodies
has been derived. Following the same line of reasoning, we generalize
that result to our case. Let us recall the  point like limit of the generally
covariant energy momentum tensor for a particle of mass $m$ and 
world line $x^{\mu}(\tau)$ \cite{Weinberg} 
\be
\tilde{T}^{\mu\nu}(x')=\frac{p^{\,\mu}p^{\,\nu}}{p^{\,0}\sqrt{-\tilde{g}}}\,\delta^{(3)}
\bigl(x'-x(\tau)\bigr),
\ee
where $p^{\mu}=m d x^{\mu}/d \tau$. We can rewrite the scalar charge density
 $\tilde{\sigma}_{a}$ for a test body, in terms of dimensionless
scalar functions ${\tilde{q}}_{a}$, which express the 
relative strengths of non universal scalar to tensor forces 
\be \label{scalarcharges}
\tilde{\sigma}_{a}(x')=-\tilde{q}_a \tilde{T}(x')=
\tilde{q}_{a}\,\frac{m^2}{p^{\,0}\sqrt{-\tilde{g}}}\;
\delta^{(3)}\bigl(x'-x(\tau)\bigr).
\ee
As we consider long range fields, $\tilde{q}_{a}\ll 1$ to 
avoid conflicts with the present test of the Weak Equivalence Principle.
From (\ref{conservation}) we get the geodesic equation in 
scalar tensor theory with non minimal couplings \cite{g1}
\be
\label{worldline}\ddot{x}^{\,\mu}+\,\tilde{\Gamma}^{\,\mu}_{\,\alpha\nu}\,\dot{x}^{\,\alpha}
\dot{x}^{\,\nu}+\tilde{q}_{a}\,\de^{\,\mu}\varphi^{a}=\,0\;,
\ee  
where $\dot{x}^{\,\mu}\equiv {\d}x^{\,\mu}/{\d}\tau$. Now we can compute 
the modifications to the relative acceleration between two test mass bodies 
moving along two worldlines induced by the  
$\tilde{q}_{a}$'s.

Let us take two weakly self gravitating bodies moving along two infinitesimally 
close worldlines $x^{\mu}(\tau)$ and \hbox{$x'^{\mu}(\tau)=x^{\mu}(\tau)+
\delta^{\mu}(\tau)$}, 
where $\delta^{\mu}$ is the separation vector between the two curves. If we suppose that the
bodies have different scalar couplings $\tilde{q}^{(1)}_{a}$ and $\tilde{q}^{(2)}_{a}$, 
their relative acceleration is \cite{Weinberg}
\be\label{acceleration}
\ddot{\delta}_i=-\,\Bigl[\tilde{R}_{iojo}
+\tilde{q}^{(2)}_{a}\de_i\de_j\varphi^{a}\Bigr]\,\delta_j+\,\Bigl[\tilde{q}^{(1)}_{a}-
\tilde{q}^{(2)}_{a}\Bigr]\de_i\varphi^a\;,
\ee
where $ \dot{\delta}_{\,i}\equiv{\d}\delta_{\,i}/{\d} t$. Notice that in
(\ref{acceleration}) there is a term proportional to 
$\tilde{q}^{(1)}_{a}-\tilde{q}^{(2)}_{a}$.
This term will be important when the test mass bodies
are of different nature (e.g. one 
is a baryon and the other one a lepton) but it is irrelevant inside
a GW detector.
Therefore, the equation needed to analyze the response of \text{GW} 
detectors to scalar tensor waves is
\be\label{response}
\ddot{\delta}_i=-\,\Bigl[\tilde{R}_{iojo}+\tilde{q}_{a}
\de_i\de_j\varphi^{a}\Bigr]\,\delta_j\;.
\ee

\subsection{Gravitational waves}\label{secGW}

Let us recall some results concerning scalar tensor GWs \cite{Damour} . 
In the weak field limit of the theory 
\bea
\tilde{g}_{\mu\nu} (x) &=&\eta_{\mu\nu} + \tilde{h}_{\mu\nu} (x),\nonumber \\ 
\varphi^{a} (x) &=& \varphi^a_{0} + \xi^a (x),
\label{verylast}\eea
where $\vert \tilde{h}_{\mu\nu}\vert \ll 1, \;
\vert\xi^a \vert\ll 1$, $\eta_{\mu\nu}$ is the flat Minkowski metric, and 
$\varphi_0^a$ the background values of the scalar fields.
We now choose a gauge in which the metric perturbation 
has zero time-time and time-space components while the purely spatial 
components, for a plane wave propagating along the  direction
characterized by the unit vector $\O$, assume 
the form 
\be\label{wave1}
\tilde{h}_{ij}(x)\,=\,h_{A}(x)\,e^{A}_{ij} (\O) + 
2 \alpha_{a}^{0}\,\xi^{a}(x)\,e^{s}_{ij} (\O); \qquad \qquad 
A = +,\times;\quad a = 1,\ldots,n.
\ee
$e^+$, $e^{\times}$ are the spin 2 polarization tensors describing the 
ordinary GW in the transverse traceless gauge, $e^{s}$ is the spin 0 
polarization tensor of the scalar waves, 
$\alpha_a^0\equiv\alpha_a(\varphi^a_0)$, and we choose units such that 
$A(\varphi_0^a)=1$. By indicating with $\m$ and $\n$ a pair of orthonormal 
vectors lying in the plane perpendicular to $\O$, these polarization 
tensors can be written as follows (see Appendix \ref{appendixC2}) 
\bea\label{polarization}
e^{+}_{ij}(\O) &=& \m_{i}\m_{j} - \n_{i} \n_{j}, \nonumber \\
e^{\times}_{ij}(\O) &=&\m_{i}\n_{j} + \n_{i} \m_{j}, \\
e^s_{ij}(\O) &=& \delta_{ij} - \O_i \O_j =
\m_{i}\m_{j} + \n_{i} \n_{j},\nonumber
\eea 
and 
$$
e^{B}_{ij}(\O)\,e^{B'ij}(\O)=2 \delta^{B B'} \qquad \qquad
B = +,\times,{s}\;.
$$

We consider now the small relative oscillations of two weakly self gravitating 
bodies induced by this wave. By indicating with $L_i$ the rest separation of
the bodies, we can put $\delta_{i} = L_i + \zeta_i$ ($\zeta_i \ll 1$). 
Expanding (\ref{response}) to first order in $\zeta_i$, we find 
\be
\ddot{\zeta}_i \,=\, -\frac12\,\Biggl[\frac{{\d}^2 \tilde{h}_{ij}}
{{\d} t^2} + 2\tilde{q}_{a}\de_i\de_j\xi^{a}\Biggr]L_j.
\ee
Since we are considering plane wave solutions, the spatial derivatives appearing 
in the last equation can be replaced by the time derivatives, namely 
$\de_{i} \de_{j} \xi^a = \O_i \O_j \ddot{\xi}^a=
(\delta_{ij}-e_{ij}^{s}(\O))\ddot{\xi}^a$, 
and taking into account (\ref{wave1}), one finds  
\be
\ddot{\zeta}_i \,=\, -\frac12\,\frac{{\d}^2}{{\d} t^2}\Biggl[h_{A}(x)\, 
e^{A}_{ij}(\O) + 2 \left(\alpha_a^0\,
- \tilde{q}_a\right)\xi^a(x)e_{ij}^{s}(\O)+2\tilde{q}_a\xi^a(x)\delta_{ij} 
\Biggr] L_j,
\ee
and then the infinitesimal displacement induced by the GW is  
\be\label{wavenot}
\zeta_i \,=\, -\frac12\Biggl[h_{A}(x)\, 
e^{A}_{ij}(\O) + 2 \left(\alpha_a^0\,
- \tilde{q}_a\right)\xi^{a} (x)e_{ij}^{s}(\O)+2\tilde{q}_a\xi^a(x)\delta_{ij} 
\Biggr] L_j.
\ee
This formula needs a few comments. The scalar fields considered in our theory 
are massless, therefore the scalar GW can carry energy and momentum through just
one degree of freedom, the {\it transverse} polarization tensor $e_{ij}^s(\O)$ 
(see Wagoner in Ref. \cite{Brans}). 
Therefore in (\ref{wavenot}) only the transverse part strains the matter and the 
$\delta_{ij}$ is effectively unimportant when studying the response the antennas 
to GWs. By introducing the {\it effective} gravitational wave sensed by the 
test mass bodies 
\be
\tilde{h}_{ij}^{eff}=h_{A}(x)\, e^{A}_{ij}(\O) + 2 
\left(\alpha_a^0\,
- \tilde{q}_a\right)\xi^{a} (x)e_{ij}^{s}(\O)\;
\ee
we rewrite (\ref{wavenot}) as follows
\be\label{bobetta}
\zeta_i \,=\, -\frac12 \tilde{h}_{ij}^{eff} L_j\;.
\ee
However, 
if the scalar fields were slightly massive, there would be also a 
{\it longitudinal} polarization along the propagation direction of 
the GW and we couldn't drop the $\delta_{ij}$ in (\ref{wavenot}).
This scenario has been analyzed in \cite{maggiore,g1,g2,g3}, but 
in the following we won't consider it and just restrict our study to 
$\tilde{h}_{ij}^{eff}$.

\section{Interferometers and resonant mass spherical detectors}

\subsection{Response function of an interferometer to scalar GWs}
Let us consider a Michelson type laser interferometer with two orthogonal 
arms of the same nominal length $L_1=L_2=L$. From (\ref{bobetta}), the 
signal at the output port of the interferometer (the strain of the 
differential mode) is proportional to the difference in the two path 
lengths, $\zeta_1-\zeta_2$, induced by the wave and can be written in the 
form \cite{Forward} 
\be
\tilde{h}^{eff}=\tilde h_{ij}^{eff}{\cal D}^{ij},
\label{interf}
\ee
where ${\cal D}$ is a traceless and symmetric tensor describing the geometry of the  interferometer  
\footnote{(\ref{interf}) is valid in the regime in which the wavelength
of the impinging scalar GW is much bigger than the length of the arms of 
the interferometer. Given the resonant frequencies of our resonant mass 
detectors, this will be always the case in the present paper. For a more 
detailed discussion of this point see \cite{nakao}.}. 
In the interferometer frame, 
namely the one where the corner station stands at the origin of coordinates and the
$\x$ and $\y$ axes lie along the arms, this tensor writes 
\be\label{Michelagnolo}
{\D}=\frac{1}{2}
\pmatrix{
1&0&0\cr
0&-1&0\cr
0&0&0}.
\ee
The effective strain sensed by the interferometer is then split in a spin 2 
and a spin 0 part, proportional to the difference $\alpha_a^0-\tilde{q}_a$; we 
can take explicitly into account the dependence of the strain from the angles 
$(\theta,\phi)$ defining the direction $\O$ of the incoming wave by introducing 
the angular pattern functions of the interferometer  
\be\label{patti}
F^{A}(\O) = e_{ij}^{A}(\O){\cal D}^{ij},\qquad\qquad 
F^{s}(\O) = e_{ij}^{s}(\O){\cal D}^{ij},
\ee
and writing the strain as
\be
\tilde{h}^{eff}=h_{A}(x)\, F^{A}(\O) + 2 \left(\alpha_a^0\,
- \tilde{q}_a\right)\xi^{a} (x)\,F^{s}(\O)\;.
\ee

\subsection{Cross section for resonant spheres in scalar tensor theory}

We discuss now the cross section of a resonant sphere in the general scalar 
tensor theory. For spin 2 waves this result was obtained in 
Ref. \cite{WagonerPaik} (see also Ref. \cite{Zhou}). 
In recent years, this kind of detector (both solid and hollow) 
has been extensively studied as device able to analyze the 
spin content of GWs (see Refs. \cite{Lobo,Loboortega,Colacino}). 
The calculation of its scattering cross section in the framework of 
the Brans-Dicke theory was carried out in Refs. \cite{Brunetti,Fucito,Cocciafucito}.
The extension of the results of Ref. \cite{Brunetti} to 
the general scalar tensor theory with minimal coupling is
straightforward \cite{Luca} and in the following subsections 1 and 2, we will 
just sketch the steps and quote the results.
Furthermore, in subsection 3 we will repeat the calculations for the even
more general case of $\tilde q_{a}\neq 0$.
For the sake of generality the direction
of propagation of the wave and the  antenna frame (defined by a triad of 
orthonormal vectors $(\x,\y,\hat{z})$) will be taken to be distinct. 
The direction $\O=(\theta,\phi)$ of the incoming wave is identified by 
the relative orientation of the triad defined in (\ref{polarization})  
with respect to $(\x,\y,\hat{z})$ 
\bea\label{rotaziuncella}
\m&=&\cos\phi\,\x + \sin\phi\,\y ,\nonumber\\
\n&=&-\sin\phi\cos\theta\,\x + \cos\phi\cos\theta\,\y + \sin\theta\,\hat{z},\\
\O&=&\sin\phi\sin\theta\,\x  -\cos\phi\sin\theta\,\y   + \cos\theta\,\hat{z}.
\nonumber
\eea

\subsubsection{Tensor GWs}

Consider a superposition of spin 2 plane GWs with wave vector $k^{\mu}$
and amplitudes $h_{A}$ impinging on a spherical GWs detector 
\be
\label{scattering}
\tilde{h}_{\mu\nu}\equiv
\tilde{e}_{\mu\nu}e^{ik_{\rho}x^{\rho}}+ c.c.\equiv h_{A}e^{A}_{\mu\nu}
e^{ik_{\rho}x^{\rho}}+c.c.,
\qquad A=+,\times.
\ee
Note that hereafter $e^{A}_{\mu\nu}$ are the
polarization tensors written in the detector frame
$(\x,\y,\hat{z})$ (see Appendix \ref{appendixC2} for their explicit expressions).
As usual we will use the so called quadrupole
approximation, i.e. we suppose that the detector is much
smaller than the wavelength of the impinging GW, so that
only the first terms (quadrupole, for the tensor
component; monopole and quadrupole for the scalar one)
have to be considered. Analogously to Ref. \cite{Weinberg} we
find the expressions for the spin 2 scattering and total energy
cross sections 
\bea \label{sez urto 1}
\sigma^{scat}_{h}&=&\frac{128\pi  G^2}{5}\,\,\frac{\Bigl[1+
\frac{1}{3}\,\alpha_{0}^2\left(1-\alpha_0^2\right)\Bigr]
\tilde{\tau}^*_{ij}\tilde{\tau}^{ij}}{\tilde{e}^*_{ij}\tilde{e}^{ij}}\\
\sigma^{tot}_{h}&=&\frac{8  G}{f}
\frac{\Im(\tilde{e}^*_{ij}\tilde{\tau}^{ij})}{\tilde{e}^*_{ij}\tilde{e}^{ij}},
\eea
where $\alpha_0^2=\alpha^0_a\alpha_0^a$ and
$\tilde{\tau}_{ij}\equiv \tilde{\tau}_{ij}(\O,f)$ is the (traceless) 
Fourier transform of 
the variation induced in the stress energy tensor of the
sphere by the impinging GW\footnote{In principle the expression of $\tilde{\tau}_{ij}$ could
contain also a term proportional to
$\D^{(00)}_{ij}\propto\delta_{ij}$ \cite{Weinberg}, accounting for the trace of the
polarization tensor (monopole excitation). But, since the trace
$\tilde e_{ii}$ vanishes, in this tensorial part such a term
gives no contribution.}. 
Furthermore we will study resonant scattering,
i.e. we will assume that the detector scatters only the impinging GWs with
frequency $f$ around the resonant frequency of one of its
natural vibrational modes. This leads to a relation between
$\sigma^{scat}_{h}$ and $\sigma^{tot}_{h}$ and to
another between $\tilde{\tau}_{ij}$ and the sphere mode tensors 
\bea \label{sigma eta}
\sigma^{scat}_{h}=\eta\sigma^{tot}_{h},\\
\label{tauS}
\tilde{\tau}_{ij}=\gamma(f)\tilde{e}_{ij},
\eea
where $\eta$ is the fraction of the total oscillation energy dissipated through 
emission of GWs (it can be calculated as a function of the detector internal 
parameters) and $\gamma(f)$ gives the frequency dependence of $\tilde{\tau}_{ij}$. 
The function $\gamma(f)$ is chosen so that the response of the antenna is 
resonant in frequency (see Refs. \cite{Lobo,Brunetti}). If the oscillation of
the mode with angular momentum $l=2$ has proper 
frequency\footnote{The index $n=1,\ldots,\infty$ labels
different solutions for the sphere eigenmodes with fixed
angular momentum  $l$ \cite{Brunetti}.}
$f_{n2}$ and a bandwidth $\Delta_{f_{n2}}$, we find 
\be \label{propto}
\gamma(f) \propto \frac1{f - f_{n2} + i\Delta_{f_{n2}}/2}.
\ee
Substituting (\ref{sez urto 1}) and (\ref{propto})
respectively into (\ref{sigma eta}) and (\ref{tauS}) and
combining the results, the total energy cross section becomes 
\be \label{tscatt}
\sigma_{h}(f;\,n,l=2)\equiv\sigma^{{tot}}_{h}(f)=\frac{1}{1+\alpha_{0}^2}
\frac{\tilde{G}M v^2 F_n}{2\pi }
\frac{\Delta_{f_{n2}}}{\left(f-f_{n2}\right)^2+
\Delta_{f_{n2}}^2/4},
\ee
where $M$ is the sphere mass, $v$ the velocity of
sound in the material the sphere is made of, $F_n$ a
constant depending only on the quadrupolar mode under scrutiny and on the 
sphere parameters (radius, density, material) \cite{Brunetti}, and 
$\tilde G=(1+\alpha^2_0)G$
is the effective Newton's constant measured in Cavendish like experiments. 
The results of Ref. \cite{Brunetti} in Brans-Dicke theory are recovered by 
setting $\alpha_0^2=(2\omega_{BD}+3)^{-1}$ where $\omega_{BD}$ is the 
Brans-Dicke parameter. If in performing this calculation we expand the 
tensors in the numerator of $\sigma^{tot}_{h}$ in the detector basis
${\cal D}^{(\epsilon)}_{ij}$ (defined in Appendix \ref{appendixC1}),
we find that the total cross section can be written as
the sum of five terms,
the total cross sections for any single vibrational mode of
the sphere 
\be
\sigma^{tot}_{h}(f)=\sum_\epsilon \sigma^{(\epsilon)}_{h}(f,\O),\qquad
\qquad \epsilon=0,1c,1s,2c,2s, 
\ee
and
\be
 \sigma^{(\epsilon)}_{h}(f,\O) =\frac{1}{1+\alpha_{0}^2}
\frac{\tilde{G}M v^2 F_n}{2\pi}
\frac{\Delta_{f_{n2}}}{\left(f-f_{n2}\right)^2+
\Delta_{f_{n2}}^2/4}\,
\frac{\sum_A \left|F^{(\epsilon)}_{A}h_{A}\right|^2}
{\sum_{A'} |h_{A'}|^{\,2}}.
\label{tscatt1}\ee
The angular dependence is enclosed in the pattern functions, 
$F^{(\epsilon)}_{A}\equiv {\D}^{(\epsilon)}_{ij}e_{A}^{ij}(\O)$, 
explicitly written in Table \ref{table1}.

For later purposes we will need also the integrated cross section. 
By integrating (\ref{tscatt}) we get 
\be
\Sigma_{h}\left(n;l=2\right)=\frac{1}{1+\alpha_{0}^2}\, 
\tilde{G}M v^2F_n\;.\label{lastmom}
\ee

\subsubsection{Minimally coupled scalar GWs}
The scattering and total 
cross section for the minimally coupled scalar part of a GW are 
\bea
\sigma^{scat}_{s}&=&\frac{8\pi G ^{2}\alpha_0^2}{5}
\frac{\big[\mid\tilde{\tau}_{ii}\mid^2+\frac{1}{3}
{\tilde{\tau}}^*_{ij}\tilde{\tau}^{ij}\big]}{\xi_a^*(\vec{x},f)\xi^a(\vec{x},f)},\\
\nonumber{}\\
\sigma^{tot}_{s}&=&\frac{2 G\alpha^0_a}{f}\,\frac{\Im\big[\xi^{a}
(\vec{x},
f)e_{ij}^{s}\tilde{\tau}^{ij*}\big]}
{\xi^*_a(\vec{x},f)\xi^a(\vec{x},f)}.
\eea
where $\xi^a(\vec{x},f)$ is the Fourier transform of the impinging scalar GW.
We now have to decompose the scalar GW polarization
tensor $e_{ij}^{s}$ in a quadrupole and a
monopole part, as they excite different modes in the detector. The way to do this 
is by expressing $e_{ij}^{s}$ in the basis defined by the five 
real symmetric tensors ${\D}_{ij}^{(\epsilon)}$, plus
${\D}_{ij}^{(00)}$, proportional to the identity tensor, because 
these tensors are directly related to the angular
momentum of the excitation (see Appendix \ref{appendixC1} and Ref. \cite{Zhou}).

Assuming again resonant scattering and noting that the
resonance frequencies of the quadrupole and the monopole
modes need not be equal, we have now two expressions for
the variation of the stress energy tensor of the
detector: the first, labelled $\tilde{\tau}_{ij}(f;l=0)$ and valid 
for the sphere monopole mode, 
is proportional to ${\cal D}_{ij}^{(00)}$ and has resonance frequency 
$f= f_{n0}$; the second, $\tilde{\tau}_{ij}(f;l=2)$, is 
proportional\footnote{The $l=2$ part of $e_{ij}^{s}$ expanded in the 
${\cal D}_{ij}^{(\epsilon)}$ basis is $2\sum_\epsilon
F^{(\epsilon)}_{s}{\cal D}_{ij}^{(\epsilon)}$.} 
to $\sum_\epsilon F^{(\epsilon)}_{s}{\cal D}_{ij}^{(\epsilon)}$ and 
has resonance frequency $f= f_{n2}\neq f_{n0}$ 
\bea \label{scaligero1}
\tilde{\tau}_{ij}(f;l=0)&=&\beta'(f)\alpha_{a}^{0}\xi^{a}(f){\D}^{(00)}_{ij},\\
\label{scaligero2}
\tilde{\tau}_{ij}(f;l=2)&=&\beta''(f)\alpha_{a}^{0}\xi^{a}(f)\sum_\epsilon
F^{(\epsilon)}_{s}{\cal D}_{ij}^{(\epsilon)},
\eea
where $\beta'(f)\neq\beta''(f)$ are the analogous of the function $\gamma(f)$ 
in (\ref{tauS}) and $F_s^{(\epsilon)}\equiv{\D}_{ij}^{(\epsilon)}e^{ij}_{s}(\O)$.
We deduce then the total cross section of the monopole mode 
\be
\label{sezione00}
\sigma_{s}(f;\,n,l=0)=\frac{\alpha_{0}^2}{1+\alpha_{0}^2} 
{\tilde{G}M v^2H_n\over \pi}
\frac{\Delta_{f_{n0}}}{(f-f_{n0})^2+ \Delta_{f_{n0}}^2/4},
\ee
where $H_{n}$ is a constant depending on the monopolar mode under exam 
\cite{Brunetti} and $\Delta_{f_{n0}}$ is the resonance bandwidth. 
For the quadrupole modes ${\D}_{ij}^{(\epsilon)}$, the same calculation gives 
\be\label{sigmaesse}
\sigma_{s}(f;\,n,l=2)=\sum_\epsilon\sigma^{(\epsilon)}_{s}(f,\O)
\ee
where 
\be
\sigma^{(\epsilon)}_{s}(f,\O)=\frac{\alpha_{0}^2}{1+\alpha_{0}^2}\frac{\tilde{G}M
 v^2 F_n}{2\pi}\frac{\Delta_{f_{n2}}}{(f-f_{n2})^2+\Delta_{f_{n2}}^2/4
}\left(F^{(\epsilon)}_{s}\right)^2.
\ee
The pattern functions $F^{(\epsilon)}_s$ are listed in Table \ref{table1}: since
an explicit computation yields
\be\label{Jefferson}
\sum_{\epsilon}\left(F^{(\epsilon)}_{s}\right)^2=\frac{1}{3},
\ee
the global response to scalar waves of the quadrupole modes is isotropic too,
and total  cross section (\ref{sigmaesse}) reads
\be\label{sezione2}
\sigma_{s}(f;\,n,l=2
)=\frac{\alpha_{0}^2}{1+\alpha_{0}^2}\frac{\tilde{G}M
 v^2 F_n}{6\pi} \frac{\Delta_{f_{n2}}}{(f-f_{n2})^2+ \Delta_{f_{n2}}^2
/4}.
\label{sscatt}
\ee

As the quadrupole modes are sensitive 
to scalar and to tensor waves, the angular dependence of each cross section
could make it possible, in principle, to guess the polarization. For instance,
considering the $m=0$ mode, $F^{(0)}_{s}(\theta,\phi)$ gets a maximum 
for $\theta=\phi=0$, while  $F^{(0)}_+(0,0)=F^{(0)}_{\times}(0,0)=0$.

\begin{table}
\caption{Angular dependence of the sphere pattern functions for the three
 independent polarizations of a scalar tensor GW. Notice that the pattern functions of 
 the $\epsilon=2c$ mode coincide with the ones of the interferometer introduced in (\ref{patti})}
\begin{center}
\begin{tabular}{c c c c}
\vspace*{0.1cm}
 Mode ($\epsilon$)& $F^{(\epsilon)}_{+}(\theta,\phi)$  &
  $F^{(\epsilon)}_{\times}(\theta,\phi)$
  & $F^{(\epsilon)}_{s}(\theta,\phi)$   \\[5pt]
\hline \rule{0ex}{3.0ex}
$2s$  &  $-\cos\theta\cos 2\phi$ &
 $-\frac{1}{2}\left(1+\cos^2\theta\right)\sin 2\phi$  &
 $-\frac{1}{2}\sin 2\phi\sin^2\theta$  \\ 
$2c$  & $-\cos\theta\sin 2\phi$ &
 $\frac{1}{4}\left(3+\cos 2\theta\right)\cos 2\phi$   &
 $\frac{1}{2}\cos 2\phi\sin^2\theta$\\
$1s$ & $-\sin\theta\sin\phi$ & $\frac{1}{2}\sin
 2\theta\cos\phi$  &
 $-\frac{1}{2}\cos\phi\sin 2\theta$ \\
$1c$ & $-\sin\theta\cos\phi$   & $-\frac{1}{2}\sin
2\theta\sin\phi$ & $\frac{1}{2}\sin\phi\sin 2\theta$ 
 \\
$0$ & $0$  & ${\sqrt{3}\over 2}\sin^2\theta$ &
$\frac{\sqrt{3}}{6}(3\cos^2\theta-1)$ \\[5pt]
\end{tabular}
\end{center}
\label{table1}
\end{table}

The integration of (\ref{sezione00}) and 
(\ref{sezione2}) gives, respectively 
\be\label{rew}
\Sigma_{s}\left(n;\,l=0\right)=\frac{2\alpha_{0}^2}{1+
\alpha_{0}^2}\, 
\tilde{G}M v^2H_n,
\ee
and 
\be\label{fwd}
\Sigma_{s}\left(n;\,l=2\right)=\frac{\alpha_{0}^2}{1+
\alpha_{0}^2}\, 
\frac{\tilde{G}M v^2F_n}{3}.
\ee

\subsubsection{Non minimally coupled scalar GWs: $\tilde{q}_a\neq 0$}

In this section we present the full generalization of the result presented 
before to  the case in which $\tilde{q}_{a}$ is small but not exactly null. 
We will follow step by step the procedure outlined in Ref. \cite{Brunetti}. 
Further details for the general multi scalar metric theory can be found in 
Ref. \cite{Luca}.

\paragraph{The energy momentum conservation law\\}

First let us consider the energy momentum tensor $\tilde{T}^{\mu\nu}$ of the 
resonant sphere
and write the linearized conservation law (\ref{conservation}) in momentum 
space. Denoting by
$\tilde{\tau}^{\mu\nu}(x), \tilde{\tau}(x)$ 
the linear part of $\tilde{T}^{\mu\nu}, \tilde{T}$ 
we get 
\be
\de_{\mu}\tilde{\tau}^{\mu\nu}(x)+\tilde{\sigma}_a(x)
\de^{\nu}\xi^a(x)=0,
\ee
which in momentum space reads 
\be\label{stresscons}
k_{\mu}\tilde{\tau}^{\mu\nu}(k)+\tilde{\sigma}_{a}(k)*
\left[k^{\nu}\xi^{a}(k)\right]=0.
\ee
The reality of $\tilde{\tau}^{\mu\nu}(x)$ and $\xi^a(x)$ implies
$\tilde{\tau}^{\mu\nu*}(k)=\tilde{\tau}^{\mu\nu}(-k)$ and $\xi^{a*}(k)=
\xi^a(-k)$, with $k=(\vec k, \omega)$ and $\omega\equiv k_{0}=|\vec{k}|$.
The asterisk in (\ref{stresscons}) stands for the four dimensional convolution 
product. Now we proceed to express $\tilde{\tau}_{00}(k)$ in terms of 
$\tilde{\tau}_{ij}(k)$.

For a particle of mass $m$, (\ref{scalarcharges}) defines the relation between 
the scalar charge densities and the components of the energy momentum tensors. 
Integration over all particles of the resonant sphere gives 
\be
\tilde{\sigma}_{a}(x)=-\tilde{q}_{a}\tilde{\tau}(x),
\ee
and therefore the four equations (\ref{stresscons}) in momentum space read
 (with \mbox{$\tilde{\tau}_{\mu\nu}(k)\equiv \tilde{\tau}_{\mu\nu}$}) 
\bea
\label{fleur1}
k_{0}\tilde{\tau}^{00}+k_i\tilde{\tau}^{0i}-\tilde{q}_a \tilde{\tau}
*\left[k^0 \xi^a(k)\right]=0,\\
\label{fleur2}
k_{0}\tilde{\tau}^{0i}+k_j\tilde{\tau}^{ij}-\tilde{q}_a 
\tilde{\tau}
*\left[k^i \xi^a(k)\right]=0.
\eea
The wave travels along the direction $\O$, and so $k_i=k_0\O_i$ because the 
scalar fields are massless. Subtracting the contraction of (\ref{fleur2}) 
with $\O_i$ from (\ref{fleur1}) gives then 
\be\label{linear}
\tilde{\tau}_{00}=\tilde{\tau}_{ij}\O^i\O^j\;, 
\ee
a relation which holds in minimally coupled scalar tensor theories too \cite{Luca}.

We now compute again, in the case $\tilde{q}_a\neq 0$, the quantities entering 
the cross sections: the incoming energy flux, the power emitted by the detector 
in GWs and the interference power (see Ref. \cite{Weinberg}). The calculation 
strictly follows that of Ref. \cite{Brunetti} for Brans-Dicke theory which has 
been generalized in Ref. \cite{Luca} to multi scalar metric theory. These latter 
results are recovered in the limit $\tilde{q}_a\to 0$.

\paragraph{The incoming energy flux\\}

Let us start with the incoming energy flux which
is independent from the direction of the
incoming GW thanks to the symmetry of the detector.
We will simplify here the calculations
assuming the incoming direction to be coincident with the
$\hat{z}$ axis of the detector frame. Later we will
recover the general expression for an arbitrary direction.
 
The incoming flux is computed given the energy momentum
pseudotensor of the gravitational field.
At second order in the linear expansion, defined in (\ref{verylast})
\bea\label{def di pseudotensore}
\tilde{t}^{(2)}_{\mu \nu}&=&\frac{2}{8\pi G}\bigg\{-(\de_a\alpha_b)_0\xi^a
\de_\mu\de_\nu\xi^b+\frac{1}{2}\alpha_b^0\eta^{\rho\sigma}(\de_\mu \tilde{h}_{\nu\rho}+
\de_\nu \tilde{h}_{\mu\rho}-\de_\rho \tilde{h}_{\mu\nu})\de_\sigma\xi^b+\nonumber\\
&&-(\de_a\alpha_b)_0\eta_{\mu\nu}\eta^{\rho\sigma}\xi^a\de_\rho\de_\sigma\xi^b+
\frac{1}{2}\alpha_b^0\eta_{\mu\nu}\eta^{\epsilon\gamma}\eta^{\rho\sigma}(\de_\epsilon 
\tilde{h}_{\gamma\rho}+\de_\gamma \tilde{h}_{\epsilon\rho}-\de_\rho \tilde{h}_{\gamma
\epsilon})\de_\sigma\xi^b+\nonumber\\
&&-\big[(\de_a\alpha_b)_0+\alpha_a^0\alpha_b^0-\gamma_{ab}^0\big]\de_\mu\xi^a\de_\nu\xi^b-
\bigg[(\de_a\alpha_b)_0-\frac{1}{2}\alpha_a^0\alpha_b^0-\frac{1}{2}\gamma_{ab}^0\bigg]
\eta_{\mu\nu}\eta^{\rho\sigma}\de_\rho\xi^a\de_\sigma\xi^b\bigg\}+\nonumber\\
&&-\frac{1}{8\pi G}\bigg(\tilde{R}^{(2)}_{\mu \nu}-
\frac{1}{2}\eta_{\mu \nu}\eta^{\alpha
\beta}\tilde{R}^{(2)}_{\alpha \beta}+
\frac{1}{2}\eta_{\mu \nu}\tilde{h}^{\alpha \beta}\tilde{R}^{(1)}_{\alpha \beta}
-\frac{1}{2}\tilde{h}_{\mu \nu}\eta^{\alpha \beta}\tilde{R}^{(1)}_{\alpha \beta}\bigg),
\eea
where
\bea
\tilde{R}^{(2)}_{\mu \nu}&=&\frac{1}{2}\tilde{h}^{\alpha \rho}\bigg(\partial_\mu 
\partial_\nu \tilde{h}_{\alpha \rho}-\partial_\mu \partial_\rho \tilde{h}_{\alpha \nu}-
\partial_\mu \partial_\alpha \tilde{h}_{\nu \rho}+\partial_\alpha \partial_\rho 
\tilde{h}_{\mu \nu}\bigg)+\nonumber\\
&&+\frac{1}{4}\bigg(\partial_\mu \tilde{h}_{\alpha \rho}+
\partial_\alpha \tilde{h}_{\mu \rho}-\partial_\rho \tilde{h}_{\alpha \mu}\bigg)
\bigg(\partial_\alpha \tilde{h}^\rho_\nu+\partial_\nu \tilde{h}^{\alpha \rho}-
\partial_\rho \tilde{h}^\alpha_\nu \bigg)+\nonumber\\
&&-\frac{1}{4}\bigg(\partial_\mu \tilde{h}_{\nu \rho}+\partial_\nu \tilde{h}_{\mu \rho}-
\partial_\rho \tilde{h}_{\mu \nu}\bigg)\bigg(2\partial_\alpha \tilde{h}^{\alpha \rho}-
\partial^\rho \tilde{h}\bigg),
\eea
is the Ricci tensor linearized to second order in the fields. We keep in mind 
that \cite{Luca}
\bea\label{Xanto}
\tilde{h}_{\mu\nu}=\pmatrix{
0&     0        &0      &0\cr
0&   {\cal E}^+ +2\alpha_a^0\xi^a   &   {\cal E}^\times&0\cr
0&   {\cal E}^\times                &    -{\cal E}^+ +2\alpha_a^0\xi^a&0\cr
0&    0    &   0   &    0
},
\eea
and denote by $<\dots>$ the integration over a 3 dimensional space region 
with linear dimensions much bigger
than the GWs wavelength.
Substituting (\ref{Xanto}) into (\ref{def di pseudotensore}) 
we obtain the total scalar tensor energy flux coming from the 
$\hat{z}$ direction 
\be \label{pseudotensore1}
\Phi(f)=\Phi_h+\Phi_s=\hat{z}<\tilde{t}^{(2)}_{0z}>=
\frac{\pi f^2}{2G}\biggl\{|{\cal E}_+|^{\,2}+|{\cal E}_\times|^{\,2}+
4\gamma^0_{ab}\xi^{a*}(\vec{x},f)\xi^b(\vec{x},f)\biggr\}.
\ee

\paragraph{The scattering amplitude and the energy cross sections\\}

Let us consider a GW impinging onto our spherical resonant detector.  
At large distances, $R=|\vec{x}|$, from the detector
\be
\xi^a(\vec{x},t)\rightarrow\biggl[\xi^a(\vec{x},f)e^{i\vec{k}\cdot\vec{x}}+
\Delta^a(\vec{x},f)
\frac{e^{2\pi i f R}}{R}\biggr]e^{-2\pi i f t},
\ee
where $\Delta^a(\vec{x},f)$ is the scattering amplitude relative to the 
$a$th scalar field. It obeys the usual reality condition 
$\Delta^a(\vec{x},f)=\Delta^{a*}(\vec{x},-f)$. Using the scalar field 
equation (\ref{fieldscalar}), under the hypothesis that the quadrupole 
approximation holds, the scattering amplitude can be written in terms of 
$\tilde{\tau}^{\mu\nu}(x)$ as 
\bea\label{int per Delta}
\Delta^a(\vec{x}, \omega) &\simeq&
G\int {\d}^3 x' \left(\alpha_0^a-\tilde{q}^a\right)
\tilde{\tau}(\vec{x}',\omega) e^{-i\vec{k}\cdot\vec{x}'} \nonumber \\
&=& G\left(\alpha_{0}^a-\tilde{q}^a\right)\,\tilde{\tau}_{ij}(k)\,
\left(\delta^{ij}-\O^i\O^j\right),
\eea
where we have expressed $\tilde{\tau}_{00}(k)$ in terms of the space like 
components of the Fourier transform of the energy momentum tensor by making 
use of (\ref{linear}).

Let us turn then to the detailed calculation of the energy cross section, 
referring ourselves again to Ref. \cite{Brunetti}. From (\ref{int per Delta}) 
we find the scattering power to be 
\be
P^{scat}=\frac{2\pi f^2}{G}\int {\d} \O\,\Delta_a(\vec{x}, f)
\Delta^{a*}(\vec{x}, f) =\frac{16 \pi^2 G f^2}{5}
\biggl\{|\tilde{\tau}_{ii}|^{\,2}+\frac{1}{3}\tilde{\tau}^*_{ij}
\tilde{\tau}^{ij}\biggr\}
\left(\alpha_{0}^2-2\tilde{q}_{a}\alpha_{0}^a+\tilde{q}^2\right),
\ee
where $\tilde{q}^2=\tilde{q}_a\tilde{q}^a$. Furthermore, the interference 
between the incident plane wave and the scattered wave  gives 
\be
P^{int}=\frac{2 f}{G}\Im\,\bigg[\int {\d}\O \,\xi_a(\vec{x},f)
\Delta^{a*}(\vec{x},f) \delta(1-\hat{k}\cdot\hat{x})\bigg] = 
4\pi f\,\Im\bigg\{\xi_a^{*}(\vec{x},f)
\left(\alpha_0^a-\tilde{q}^a\right)\tilde{\tau}_{ij}e^{ij}_{s}\bigg\}.
\ee
The scattering and total cross sections are then  
\bea
\label{kugelscat}
\sigma^{scat}_{s}&=&\frac{P^{scat}}{\Phi_{s}}=\frac{8\pi
G^2}{5}\left(\alpha_{0}^2-2\tilde{q}_{a}\alpha_{0}^a+\tilde{q}^2\right)
\frac{\Bigl\{|\tilde{\tau}_{ii}|^{\,2}+\frac{1}{3}\tilde{\tau}_{ij}^*
\tilde{\tau}^{ij}\Bigr\}}
{\xi_{c}^*\,\xi^{c}},\\
\nonumber{}\\
\sigma^{tot}_{s}&=&-\frac{P^{int}}{\Phi_{s}}=
\frac{2 G}{f}\frac{\Im\Big\{\left(\alpha^0_a-\tilde{q}_a\right)\xi^{a*}
\,\tilde{\tau}_{ij}e^{ij}_{s}\Big\}}
{\xi_{c}^*\,\xi^{c}},
\eea
where we have put $\xi^c\equiv\xi^{c}(\vec{x}, f)$.
Expanding now in the ${\cal D}^{(00)}_{ij}$, ${\cal
D}^{(\epsilon)}_{ij}$ basis, we can decompose $\tilde{\tau}_{ij}$ into an
$l=0$ part and an $l=2$ part
\bea
\label{f_00}
\tilde{\tau}_{ij}(f;l=0)&=&\zeta'(f) \left(\alpha_{a}^0-\tilde{q}_{a}\right)\xi^a{\D}_{ij}^{(00)},\
\qquad \qquad\qquad\quad\quad\;\; f= f_{n0},\\
\label{f_2m}
\tilde{\tau}_{ij}(f;l=2)&=&\zeta''(f)\left(\alpha_{a}^0-\tilde{q}_{a}\right)\xi^a\sum_\epsilon 
F_{s}^{(\epsilon)}
(\O) {\cal D}_{ij}^{(\epsilon)},\qquad \qquad f= f_{n2},
\eea
with $\zeta'(f)\neq\zeta''(f)$ defined as $\beta'(f)$ and $\beta''(f)$ in (\ref{scaligero1}) and
(\ref{scaligero2}).
Hence, the monopole and the quadrupole total cross sections 
become 
\bea
\label{kugel1}
\sigma_{s}(f;n,l=0)&\equiv&\frac{2G}{f}\;\Im(\zeta')\;\frac{L\left(\xi^a;\alpha_a^0,
\tilde{q}_a\right)}
{\xi_c^*\,\xi^c},\\
\nonumber{}\\
\label{kugel2}
\sigma_{s}(f;n,l=2)&\equiv&\frac{2G}{f}\;\Im(\zeta'')\;
\sum_\epsilon{\left(F_{s}^{(\epsilon)}\right)^2}\;\frac{L\left(\xi^a;\alpha_a^0,
\tilde{q}_a\right)}
{\xi_c^*\,\xi^c},
\eea
where 
\be
L\left(\xi^a;\alpha_a^0,\tilde{q}_a\right)\equiv|\,\alpha_{a}^0\,\xi^a|^{\,2}+|\,
\tilde{q}_a\,\xi^{a}|^{\,2}-
\left(\alpha_{a}^0\,\tilde{q}_b\,\xi^{a*}\,\xi^b+\tilde{q}_a\,\alpha_b^0\,\xi^{a*}\,
\xi^{b}\right).
\ee
By using (\ref{kugelscat}) and (\ref{f_00})-(\ref{kugel2}) with the analogous of (\ref{sigma eta}) 
with $\sigma_s$ replacing $\sigma_h$, and assuming once again 
resonant 
scattering,
we get 
the final form for the monopole and quadrupole total cross sections 
\bea
\sigma_{s}(f;n,l=0)&\equiv&\frac{\eta_0}{\pi f^2 \left(\alpha_0^2-
2\tilde{q}_a\alpha_0^a+
\tilde{q}^2\right)}\;
\frac{\Delta^2_{f_{n0}}/4}{(f-f_{n0})^2+\Delta_{f_{n0}}^2/4}\;
\frac{L\left(\xi^a;\alpha_a^0,\tilde{q}_a\right)}
{\xi_c^*\,\xi^c},\\
\nonumber{}\\
\sigma_{s}(f;n,l=2)&\equiv&\frac{15\,\eta_2}{\pi f^2 \left(\alpha_0^2-2
\tilde{q}_a
\alpha_0^a+\tilde{q}^2\right)}\;
\frac{\Delta^2_{f_{n2}}/4}{(f-f_{n2})^2+\Delta_{f_{n2}}^2/4}\;
\sum_\epsilon{\left(F_{s}^{(\epsilon)}\right)^2}\;\frac{L\left(\xi^a;
\alpha_a^0,
\tilde{q}_a\right)}
{\xi_c^*\,\xi^c}.
\eea

We still have to evaluate $\eta_{0}$ and $\eta_2$. This is done remembering 
their definition as the ratio between the power $P^{scat}\equiv P ^{(n;l)}$ 
reemitted as gravitational waves by the vibrations of the sphere and the 
oscillatory energy $E_{osc}^{(n;l)}$ dissipated by the sphere itself 
\be
\eta_{0}=\frac{P^{(n;0)}}{2\pi\Delta_{f_{n0}}E_{osc}^{(n;0)}},\qquad
\eta_2=\frac{P^{(n;2)}}{2\pi\Delta_{f_{n2}}E_{osc}^{(n;2)}}.
\ee
The oscillatory energy is that evaluated in Refs. \cite{Lobo,Brunetti}, 
since it doesn't depend on $\tilde{q}_a$. The calculation of the reemitted
power follows that of Ref. \cite{Brunetti,Luca}. The only difference consists
in replacing $\alpha_a^0$ with $\alpha^0_a-\tilde{q}_a$. Therefore, omitting 
the uninteresting details of the calculation, we get 
\bea
\eta_{0}=4 G\,\frac{M v^2f^2_{n0}\,H_n}{\Delta_{f_{n0}}}
\{\alpha_0^2-2\tilde{q}_a\alpha_0^a+\tilde{q}^2\},\\
\nonumber{}\\
\eta_2=\frac{2 G}{15}\,\frac{M v^2f^2_{n2}\,F_n}{\Delta_{f_{n2}}}
\{\alpha_0^2-2\tilde{q}_a\alpha_0^a+\tilde{q}^2\}.
\eea
Finally the cross sections assume the following simple forms 
\bea
\sigma_{s}(f;n,l=0)&\equiv&\frac{G M v^2 H_n}{\pi}\;
\frac{\Delta_{f_{n0}}}{(f-f_{n0})^2+\Delta_{f_{n0}}^2/4}\;
\frac{L\left(\xi^a;\alpha_a^0,\tilde{q}_a\right)}
{\xi_c^*\,\xi^c},\label{for1}\\
\nonumber{}\\
\sigma_{s}(f;n,l=2)&\equiv&\frac{ G M v^2 F_n}{2\pi}\;
\frac{\Delta_{f_{n2}}}{(f-f_{n2})^2+\Delta_{f_{n2}}^2/4}\;
\sum_\epsilon{\left(F_{s}^{(\epsilon)}\right)^2}\;
\frac{L\left(\xi^a;\alpha_a^0,\tilde{q}_a\right)}
{\xi_c^*\,\xi^c}
\label{for2}
\eea
These expressions can be made more manageable by expanding 
$L\left(\xi^a;\alpha_a^0,\tilde{q}_a\right)$ in powers of 
$\tilde{q}_a\ll\alpha_a^0\ll 1$, an ordering relation which 
follows from the weak field limit of (\ref{heart}).
 
First, an analogous calculation to that for $\Delta^a(\vec{x},f)$ gives 
\cite{Luca} 
\be
\xi^a(\vec{x},f)=\frac{G}{R}\left(\alpha_{0}^a-\tilde{q}^a\right)
\tilde{\tau}'_{ij}e^{ij}_{s},
\ee
where now $\tau'_{ij}=\tilde{\tau}'_{ij}(k)$ is the Fourier 
transform of 
the space components of the stress energy tensor 
of the source located at a great distance $R$ from the antenna; 
$\tilde{\tau}_{ij}'$ is related to the 
Fourier transform of the variation of the quadrupole moment $Q'_{ij}(f)$ 
of the source by \cite{Weinberg} 
\be\label{qua}
Q'_{ij}(f)= -\frac{1}{2\pi^2 f^2}\,\tilde{\tau}'_{ij}.
\ee
Therefore, taking into account (\ref{qua}), an explicit evaluation gives
\be\label{numeratore}
L\left(\xi^a;\alpha_a^0,\tilde{q}_a\right)=\frac{4\pi^4 f^4 G^2}{R^2}
|Q'_{ij}e^{ij}_{s}|^{\,2}\,
\alpha_{0}^4\,\Biggl\{1-4\frac{\tilde{q}_a\alpha_0^a}{\alpha_0^2}+2 \frac{\tilde{q}^2}
{\alpha_0^2}
+4\frac{\left(\tilde{q}_a\alpha^{a}_0\right)^2}{\alpha_0^4}
-4 \frac{\tilde{q}_a\alpha^{a}_0 \tilde{q}^2}{\alpha_0^4}+\frac{\tilde{q}^4}{\alpha_0^4}\Biggr\},
\ee
and
\be\label{denominatore}
\xi_c^*\,\xi^c=\frac{4\pi^4 f^4 G^2}{R^2}
|Q'_{ij}e^{ij}_{s}|^{\,2}\,
\alpha_0^2\,\Biggl\{1-2\,\frac{\tilde{q}_a\alpha^a_0}{\alpha_0^2}+\frac{\tilde{q}^2}
{\alpha_0^2}\Biggr\}.
\ee
Expanding this ratio in powers of $\tilde{q}_a$ yields 
\be\label{rapportino}
\frac{L\left(\xi^a;\alpha_a^0,\tilde{q}_a\right)}{\xi_a^*\,\xi^a}
\;\simeq\; \alpha_{0}^2-2\,\tilde{q}_a\alpha_0^a+\tilde{q}^2+\dots
\ee
We can finally compute the $\tilde{q}_a$ dependent terms in 
the cross sections.

The monopole cross section at the lowest order 
in $\tilde{q}_a$ reads 
\be\label{quacchero1}
\sigma_{s}(f;n,l=0)\simeq\frac{1}{1+\alpha_0^2}\,\frac{ \tilde{G} M v^2 H_n}{\pi}\;
\frac{\Delta_{f_{n0}}}{(f-f_{n0})^2+\Delta_{f_{n0}}^2/4}\;
\left\{ \alpha_{0}^2-2\,\tilde{q}_a\alpha_0^a\right\},
\ee
where we have reintroduced the effective Newton's gravitational constant $\tilde{G}$.

Analogously, the quadrupole cross section for any mode $\epsilon$ writes,
at first order 
\be
\sigma^{(\epsilon)}_{s}(f,\O)\simeq\frac{1}{1+\alpha_0^2}\frac{ \tilde{G} M v^2 F_n}{2\pi}\;
\frac{\Delta_{f_{n2}}}{(f-f_{n2})^2+\Delta_{f_{n2}}^2/4}\;
\left\{ \alpha_{0}^2-2\,\tilde{q}_a\alpha_0^a\right\}\left(F_{s}^{(\epsilon)}\right)^2.
\ee
Summing over $\epsilon$,
(\ref{Jefferson}) gives 
\be\label{quacchero2}
\sigma_{s}(f;n,l=2)\simeq\frac{1}{1+\alpha_0^2}\frac{ \tilde{G} M v^2 F_n}{6\pi}\;
\frac{\Delta_{f_{n2}}}{(f-f_{n2})^2+\Delta_{f_{n2}}^2/4}\;
\left\{ \alpha_{0}^2-2\,\tilde{q}_a\alpha_0^a\right\}.
\ee

\section{Detection of a stochastic \text{GW} background}

Our aim is to generalize the standard analysis about 
the detectability of the spin 2 stochastic 
\text{GW} background \cite{Michelson,Flanagan,Allen} to the 
case of the general scalar tensor theory outlined in 
Sec. \ref{secGW}. Within this framework we introduce a density 
of scalar gravitational radiation $\rho_s$ 
in addition to the standard tensor one $\rho_{h}$.  
If we assume, as in the tensor case, that the scalar background 
is isotropic, unpolarized, stationary and Gaussian, it is 
completely described in terms of the (dimensionless) spectrum 
\be
\Omega_s \,=\, \frac1{\rho_{c}}\frac{{\d}\rho_s}{{\d}\ln f}\;,
\ee
where ${\d}\rho_s$ is the energy density of the scalar gravitational 
radiation in the frequency range $f$ to $f+{\d}f$ and $\rho_{c}$ 
is the critical density required (today) to close the universe 
\be
\rho_{c} \,=\, \frac{3H_0^2}{8\pi \tilde{G}}\;.
\ee
$H_0$ is the present value of the Hubble constant. 
Notice that, although we study a scalar tensor theory we normalize the 
scalar gravitational spectrum to the value of $\rho_c$ recovered in general 
relativity. This choice has been taken to have a direct comparison between 
the tensor only and the scalar tensor framework. The present value of the Hubble 
expansion rate is usually written as $H_0 = h_0 \times 100$ 
km s$^{-1}$ MPc$^{-1}$, where $h_0$ (= 0.6 $\div$ 0.7) is a dimensionless 
factor that parametrizes the experimental uncertainty affecting the value of 
$H_0$. As a consequence of this definition the quantity 
$h_{0}^2 \Omega_s(f)$ is independent of $h_0$, and, thus, more
suitable to characterize the stochastic \text{GW} background.

\subsection{The signal to noise ratio for scalar tensor \text{GW} stochastic background}

From the experimental side, the signal induced in the detector output 
by a stochastic \text{GW} background is indistinguishable from the intrinsic 
noise of the detector itself. Unless the amplitude of the signal is very 
large, then, the subtraction of an a priori estimate of the detector noise 
cannot be confidently applied to the data. This implies that 
in order to detect a stochastic \text{GW} background, we should rather  analyze  
the correlated fluctuations of the outputs of, at least, two detectors 
with no common sources of noise (a condition usually verified for widely 
separated detector sites). The cross correlation among detectors 
is advantageous also from the point of view of the minimum detectable 
signal. It can be shown \cite{Flanagan,maggiore2} that, under the same 
experimental conditions, the minimum detectable signal in the correlation 
of two detectors can be even three orders of magnitude smaller than the one 
detectable with a single detector.

The problem of the optimal processing of the the detector outputs for 
the detection of the stochastic \text{GW} background (tensor and scalar) 
has been considered by various authors \cite{maggiore,Michelson,Flanagan,Allen}, 
and extensively reviewed in Ref. \cite{maggiore2}. 
This analysis can be generalized with minor modifications to the case 
of the general scalar tensor theory considered here.

The signal present at the output of each detector can be written as 
(we consider the case of two detectors) 
\be
s_k (t)\,=\,n_k (t) + \tilde{h}^{eff}_k (t),
\ee  
where we have indicated with $\tilde{h}_k^{eff}$ the gravitational strain due 
to the stochastic \text{GW} background and with $n$ the intrinsic noise of the 
detector, while $k=1,2$ labels the detector to which each quantity
is referred. The noise is assumed to be stationary, Gaussian and 
statistically independent on the gravitational strain. Furthermore, 
the assumption that the noises in the two detectors are uncorrelated 
implies that the ensemble average of their Fourier components satisfies 
\be
< n^*_k (f)\,n_l (f') >\,=\,\delta (f - f')\,\delta_{kl}\,\frac12 
S_n^{(k)} (|f|),
\ee
where $S_n^{(k)} (|f|)$ 
is the (one sided) noise power spectrum for the 
$k$th detector. 
Given an observation time $T$, the correlation 
``signal'' is defined as follows 
\be\label{sigcor}
S \,=\, \int_{-T/2}^{T/2} {\d}t\,
\int_{-T/2}^{T/2} {\d} t'\,s_1(t) s_2(t') Q(t - t'),
\ee
where $Q$ is a real filter function\footnote{This function depends 
only on $t - t'$ as consequence of the assumed stationarity of both 
the gravitational strain and the detector noise.} that, for any 
form of the signal, is chosen in order to maximize the signal to noise 
ratio (SNR) associated to $S$ \cite{Allen}.

The statistical treatment of the signal $S$ defined in (\ref{sigcor}) 
starts with the plane wave expansion of the metric perturbations. 
The effective GW exciting the detector at position $\vec{x}$ writes   
\be\label{ksignal}
\tilde{h}^{eff}_{ij} (t,\vec{x}) \,=\,\int_{-\infty}^{\,\infty}\,
{\d} f\,\int_{S^2}\, {\d} \O\,
\Biggl[h_{A}(f,\O) e^{A}_{ij}(\O)\,+\,
2\left(\alpha_{a}^{0}-
\tilde{q}_{a}\right)\xi^{a}(f,\O)e^{s}_{ij}(\O)
\Biggr]e^{2\pi i f(t-\O\cdot\vec{x})},
\ee
where, as a consequence of the reality of $\tilde{h}_{ij}^{eff} (t,\vec{x})$, 
we have that $h^*_{A} (f) = h_A (-f)$ and $\xi^{a*} (f) = \xi^a (-f)$; $\O$ is 
the unit vector specifying the direction of the incoming \text{GW}.

In terms of the detector tensor, the \text{GW} strain sensed by
the detector $k$ located at $\vec{x}_k$ is 
given by 
\be
\tilde{h}^{eff}_k (t)\,\equiv\,\tilde{h}^{eff} (t,\vec{x}_{k})\,=\,
\tilde{h}_{ij}^{eff} (t,\vec{x}_k) {\D}^{ij}_k,
\label{ksignaluno}\ee
and (\ref{ksignaluno}), keeping into account (\ref{ksignal}) and the definitions 
(\ref{patti}), it can be rewritten as
\be\label{strain}
\tilde{h}_k^{eff}(t)\,=\,\int_{-\infty}^{\,\infty}\,{\d} f\,
\int_{S^2}\,{\d} \O\,\Biggl[h_{A}(f,\O) F^A_k(\O)\,+\,
2\left(\alpha_{a}^{0} - \tilde{q}^{(k)}_{a} 
\right) \xi^{a}(f,\O) F^{s}_k(\O)\Biggr] 
e^{2\pi i f(t-\O\cdot\vec{x}_{k})}.
\ee

In the following, we focus on the spin 0 contribution to the strain, i.e. 
the part of (\ref{strain}) depending on $\xi^a(f,\O)$ (the spin 2 contribution 
has been extensively treated in Ref. \cite{Michelson,Flanagan,Allen}). 
Since the scalar background is assumed stationary, Gaussian, isotropic and 
unpolarized in the space of the scalar fields, it can be shown (see 
Appendix \ref{appendixA} for further details) that the correlation functions
between the Fourier amplitude $\xi^{a}(f,\O)$ of the waves are 
\be
< \xi^{a*}(f,\O)\,\xi^{b}(f',\O) >\,=\,
\frac1{1 + \alpha_0^2}\,\gamma^{ab}_0\,\frac{3 H_0^2}{64\pi^3}\,
\frac1{f^3}\,\Omega_{\xi} (f)\,\delta(f-f')\,\delta(\O - \O'),
\ee
where $\Omega_{\xi}(f)$ is the spectrum of a single scalar field. As a 
consequence of our assumptions on the scalar background, the whole 
spectrum is then $\Omega_{s} = n\Omega_{\xi}$. Following the same line of 
reasoning applied in the case of the spin 2 waves, under the further assumptions 
that the detector noises are much larger in amplitude than the gravitational 
strain and statistically independent on the strain itself, for the SNR we obtain 
\be\label{SNR1}
{\rm SNR}_\xi\,=\,\frac{\alpha_0^2-\left(\tilde{q}_a^{(1)}+\tilde{q}_a^{(2)}\right)\alpha_0^a+
\tilde{q}_a^{(1)}\tilde{q}_b^{(2)}\gamma^{ab}}{ 1 + \alpha_0^2}\,
\frac{3H_0^2}{8 \pi^3}\,\Biggl[2T \int_{0}^{\infty}\,{\d} f\,
\frac{\Omega_{\xi}^2 (f) \Gamma_{\xi}^2 (f)}{f^6 S^{(1)}_n(f) S^{(2)}_n(f)} 
\Biggr]^{1/2}.
\ee
The function $\Gamma_{\xi}(f)$ is the generalization to scalar fields of the usual 
overlap reduction function introduced in \cite{Flanagan,Allen}. This is a 
dimensionless function describing the reduction in sensitivity due to the different 
location and orientation of the two detectors, and it is given by 
\be\label{Gammascalar}
\Gamma_{\xi} (f) \,=\,\int_{S^2}\,{\d} \O\,F_1^{s}(\O)\,
F_2^{s}(\O)\,e^{2\pi i f d \O\cdot\hat{s}}, 
\ee
where $\hat{s}$ is the unit vector along the direction connecting 
the two detectors and $d$ is their distance.
Notice that $\Gamma_{\xi}(f)$ coincides with the scalar overlap 
reduction function introduced in Ref. \cite{maggiore} 
in the context of the single scalar metric theory of Brans-Dicke. 
The SNR
obtained in Ref. \cite{maggiore} is also recovered specializing 
(\ref{SNR1}) by setting $\tilde{q}^{(k)}_{a}=0$ and $\alpha_{0}^2=
\left(2\omega_{BD}+3\right)^{-1}$.

In the following, we will consider 
two detectors with the same 
$\tilde{q}_a^{(1)}=\tilde{q}_a^{(2)}=\tilde{q}_a$.
If we keep only linear terms in $\tilde{q}_a$, (\ref{SNR1}) becomes 
\be\label{SNR}
{\rm SNR}_\xi\,=\,\frac{\alpha_0^2-2\tilde{q}_a\alpha^a_0}{ 1 + \alpha_0^2}\,
\frac{3H_0^2}{8 \pi^3}\,\Biggl[2T \int_{0}^{\infty}\,{\d} f\,
\frac{\Omega_{\xi}^2 (f) \Gamma_{\xi}^2 (f)}{f^6 S^{(1)}_n(f) S^{(2)}_n(f)} 
\Biggr]^{1/2}.
\ee

As pointed out before, the spin 2 contribution to the strain (\ref{strain}) is 
substantially 
the one already obtained in the literature. However the presence of the scalar fields slightly 
modifies the usual formula for the SNR$_h$ of Ref. \cite{Allen} introducing
an overall $\alpha_0^2$ dependent factor which 
is absent in General Relativity 
\be\label{SNRh}
{\rm SNR}_{h}=\frac{1}{1+\alpha_0^2}\,
\frac{3H_{0}^{\,2}}{8\pi^3}\Biggl[2T\int_{0}^{\infty} {\d} f
\frac{\Omega_{h}^2(f)\Gamma^{2}_{h}(f)}{f^6 S_n^{(1)}(f)S^{(2)}_n(f)}\Biggr]^{1/2}.
\ee
Here $\Omega_{h}(f)$ is the usual spin 2 spectrum and
\be\label{Gammatensor}
\Gamma_{h}(f) \,=\,\int_{S^2}\, {\d} \O\,\sum_{A}\,F_1^{A}(\O)\,
F_2^{A}(\O)\,e^{2\pi if d\O\cdot\hat{s}},
\label{all1}
\ee 
is the (non normalized) overlap reduction function for spin 2 waves.

The most general expression for (\ref{Gammascalar}) and (\ref{Gammatensor}) 
can be shown to be \cite{Allen} 
\bea\label{overlappe}
\Gamma_w(\tau) &=& \pi\,\left\{A_w(\tau)\,\tr({\D}_1)\,\tr({\D}_2) 
\,+\,2\,B_w(\tau)\,\tr({\D}_1\,{\D}_2) \right. \nonumber \\
& & \quad +\,C_w(\tau)\,\left[{\tr}({\D}_1)\,\tr({\cal S}\,{\D}_2) 
\,+\,\tr({\D}_2)\,\tr({\cal S}\,{\D}_1)\right] \\
& & \left. \quad +\,4\,D_w(\tau)\,\tr({\cal S}\,{\D}_{1}\,{\D}_{2}) 
\,+\,E_w(\tau)\,\tr({\cal S}\,{\D}_{1})\tr({\cal S}\,{\D}_{2})
\right\} \qquad \qquad w=h,\xi \nonumber 
\eea
where $\tau = 2\pi f d$, ${\cal S}\equiv \hat{s}\otimes\hat{s}$, and
${\cal D}_k$ is the tensor of the $k$th detector. Following the procedure 
sketched in Refs. \cite{maggiore,Flanagan,Allen},
the coefficients A, B, C, D, 
and E can be expressed as linear superpositions of Bessel functions 
(see Appendix \ref{appendixB}). The traces appearing in (\ref{overlappe}) 
carry information about the geometry and the relative orientations of the 
detectors that are correlated.

\subsection{The noise power spectrum}

\paragraph{Resonant mass detectors\\} 
Let us consider a generic multi mode resonant mass antenna, with the modes 
labelled by an index $N$, as for example a resonant sphere, where $N\equiv nl$.

The noise power spectrum is a resonant curve peaked at the proper frequency 
$f_{_{N}}$ of the modes. It can be characterized by its value at the peak 
$S_n(f_{_{N}})$ and by its half height width, which gives the bandwidth of the 
resonant mode.

We now generalize the results of Refs. \cite{Loboortega,Pizzella} concerning 
$S_n(f_{_{N}})$ of resonant spheres to the scalar tensor theory. Denoting by 
$\beta_{N}$ the transducer coupling factor (the fraction of the total mode 
energy available at the transducer output), in the case of spin 2 GWs we have 
\be\label{crossh}
S_n(h;f_{_N})=\frac{1}{1+\alpha_{0}^2}\,
\frac{4 \pi\tilde{G}\hbar\beta_{_N}}{\Sigma_{_N}}=\frac{4\pi\hbar\beta_{n2}}{M v^2 F_n},
\label{diffcross}
\ee
where the expression (\ref{lastmom}) for ${\Sigma_{_N}}\equiv\Sigma_{h}(n;l=2)$ 
has been used.

For spin 0 GWs, since we have considered only the first order
terms in the expansion in powers of $\tilde{q}_a$ of the integrated cross section, 
we write 
\be\label{crossxi}
S_n(\xi; f_{_N})=\frac{\alpha_{0}^2}{1+\alpha_{0}^2}\left(1-2\frac{\tilde{q}_a\alpha_0^a}{\alpha_0^2}\right)
\,
\frac{4 \pi \tilde{G}\hbar\beta_{_N}}{\Sigma_{_N}}\; ,
\ee
where $\Sigma_N$ is now the
obvious generalization of (\ref{rew}), (\ref{fwd}) to the case $\tilde{q}_a\neq 0$. 
Making explicit (\ref{crossxi}) for both the monopole and the 
quadrupole modes, we get
\be\label{sensitivity0s}
S_n(\xi;f_{n0})=\frac{2\pi\hbar\beta_{n0}}{M v^2 H_n},
\ee
\be\label{sensitivity2s}
S_n(\xi;f_{n2})=\frac{12\pi\hbar\beta_{n2}}{M v^2 F_n}.
\ee

Finally, let us remark that formulae like (\ref{crossh}), (\ref{crossxi}) hold 
in general for any kind of detector, because the dependence of the cross 
section on the coupling constants $\tilde G$, $\alpha_a^0$ and $\tilde{q}_a$ 
doesn't change according to the geometrical features of the antenna itself. 
Actually, if no scalar fields are present, $\alpha_a^0=\tilde{q}_a=0$, 
(\ref{crossxi}) vanishes and (\ref{diffcross}) becomes the well known 
formula (4.5) of Ref. \cite{Loboortega} for the maximum sensitivity to spin 
2 waves.

The bandwidth of the resonant mode is given by 
\be
\Delta_{f_{N}}=\frac{f_{_N}}{Q_{_N}}{\Gamma_{_N}^{-1/2}},
\ee
where $Q_{_N}$ is the quality factor of the mode, which is of the order of $10^7$
and $\Gamma_{_N}$ is the ratio of the wideband noise to the narrowband noise
in the $N$th resonance mode.

\paragraph{The VIRGO interferometer\\}
In the frequency region above 2 Hz the noise power spectrum of the VIRGO 
interferometer can be approximated by the following analytical 
expression \cite{Cuoco}\footnote{With respect to this reference, the
value of $P_2$ is slightly changed as can be found 
in  http://www.virgo.infn.it/senscurve. We quote here the most recent value.
We thank the referee for pointing this out.}
\begin{equation}\label{virgospettro}
S_n(f)=P_{1}\Biggl(\frac{f_{0}}{f}\Biggr)^{5}+P_{2}\Biggl(\frac{f_0}{f}\Biggr)+
P_{3}\Biggl[1+\Biggl(\frac{f}{f_0}\Biggr)^2\Biggr],
\end{equation}
with
$$
f_0=500\,\rm{Hz}, \quad P_{1}=3.46\times 10^{-50}\,\rm{Hz}^{-1}, \quad
P_{2}=9\times 10^{-46}\,\rm{Hz}^{-1}, \quad 
P_{3}=3.24\times 10^{-46}\,\rm{Hz}^{-1}\;.
$$
In this parametrization $P_1$ and $P_2$ give the contribution of the 
pendulum and its internal modes to the thermal noise, respectively. $P_3$ 
controls instead the shot noise contribution. For frequency smaller than 
2 Hz we assume that the noise power spectrum goes to infinity.

\subsection{Sensitivity of a pair of resonant spheres}

\begin{figure}[ht]
\caption{The overlap reduction function $\Gamma_{\xi}(f)$ for two resonant spheres, 
located at relative distance $d=50$ km (solid line) and $d=400$ km (dashed line).}
\label{fig1}
\begin{center}
\includegraphics[width=200 pt, height=130 pt]{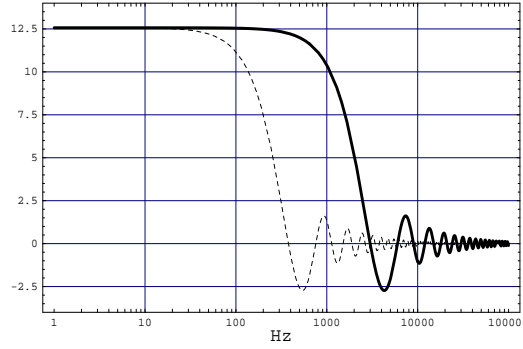}
\end{center}
\end{figure}

We consider now the correlation between two resonant spheres. 
As we are interested in scalar waves, we compute the correlation
 between the monopole modes. Since the monopole tensors are isotropic 
\be
{\D}^{1}_{ij}={\D}^{2}_{ij}={\D}^{(00)}_{ij}=\frac12\delta_{ij},
\ee
the overlap reduction function depends only on the frequency $f$ and the relative 
distance $d$. From the general formula (\ref{overlappe}) one finds 
\be
\Gamma_{\xi}(\tau)=\pi\Biggl[\frac{9}{4}A_{\xi}(\tau)+\frac{3}{2}B_{\xi}(\tau)+
\frac{3}{2}C_{\xi}(\tau)+D_{\xi}(\tau)+\frac{1}{4}E_{\xi}(\tau)\Biggr],
\ee
which explicitly reads (see Appendix \ref{appendixB}) 
\be
\Gamma_{\xi}(f)=4\pi j_{0}(\tau).
\ee 
In Fig. \ref{fig1} we plot this function for $d=50$ km, which is roughly the 
minimum distance to decorrelate seismic and e.m. noises and for $d=400$ km, 
which is the distance between the sites of the resonant bars NAUTILUS, in Frascati, 
and AURIGA in Legnaro. For $d=50$ km we observe that the first zero of the function
is around 3 kHz, which is a frequency higher than the first resonant frequency 
for both the solid mass and hollow sphere. For $d=400$ km, the first zero moves 
back at around 400 Hz. 

Restricting ourselves to metric theories ($\tilde{q}_a=0$), the SNR is
\be
\label{snrcsi}
{\rm SNR}_{\xi}=\frac{\alpha_{0}^2}{1+\alpha_{0}^2}\,\frac{3H_{0}^2}{8\pi^3}\,
\sqrt{\frac{\pi}{2}\Delta_{f_{n0}}T}
\,\frac{\Omega_{\xi}(f_{n0})\Gamma_{\xi}(f_{n0})}{f_{n0}^3 
S_n(\xi;\,f_{n0})},
\label{uno}
\ee
where $f_{n0}$ is the resonance frequency for the $n$th monopole mode. 
To evaluate (\ref{uno}) we need the 
noise power spectrum at resonance $f=f_{n0}$ given by (\ref{sensitivity0s}) where 
we assume $\beta_{n0}=0.1$ \cite{Loboortega}.
The only free parameter in 
(\ref{uno}) is $\alpha_{0}^2$ which can be
conveniently expressed in terms of the post Newtonian Eddington 
parameter \cite{Damour} as 
\be
\frac{\alpha_{0}^2}{1+\alpha_{0}^2}=\frac{1-\gamma_{\text{Edd}}}{2}.
\ee
$\gamma_{\text{Edd}}$ can be measured in light deflection experiments \cite{Lebach}. 
The minimum detectable scalar spectrum, for an observation time $T=$ 1 year, is 
found imposing SNR$_{\xi}=1$ in (\ref{snrcsi}). To isolate its dependence 
from $\gamma_{\text{Edd}}$, in Table \ref{table2} we list the reduced spectrum 
$\Omega_{\gamma_{\text{Edd}}}=\Omega_{\xi}\,(1-\gamma_{\text{Edd}})/2$.
We will comment in the Conclusions on the results obtained.

In Tables \ref{table2}, \ref{table3}, \ref{table4}, we consider hollow and solid 
mass spheres made of CuAl, Al5056 and Mo, materials which have high density and 
high velocity of sound \cite{Loboortega,Cocciafucito}. The geometrical features 
of such spheres are the outer diameter $\Phi$ and the ratio $\zeta$ between 
the inner and outer radius. We consider the first excited monopole mode, so in 
(\ref{sensitivity0s}) we put $n=1$ and $f_{10} \equiv f_0$. For solid mass 
spheres we have $H_1=1.14$ \cite{Brunetti}, while for hollow spheres $H_1$ is a 
function of $\zeta$, and some interesting values are listed in the tables 
themselves \cite{Cocciafucito}.

\begin{table}[ht]
\caption{Minimum detectable scalar spectrum (SNR$_{\xi}$ = 1, $T = 1$ year) 
for the correlation between the first monopole modes of two 
hollow spheres, with 6 meters outer diameter, made of  CuAl ($v=4700$ m/s) 
at $d=50$ km.}
\label{table2}
\begin{center}
\begin{tabular}{cccccccc} 
\vspace*{0.1cm}
      $M$ (ton)& $\zeta$ & $H_1$ & $f_{0}$ (Hz) & $\Delta_{f_{0}}$ (Hz) 
& $\Gamma_{\xi}(f_{0})$ & $\sqrt{S_n(\xi;\,f_{0})}({\rm Hz}^{-1/2})$
  &$h_{0}^2\Omega_{\gamma_{\rm Edd}}$\\[5pt]
\hline \rule{0ex}{3.0ex}
832 & 0.25 &0.73727  &770 &24.3  & 11.2  & $2.21\times 10^{-24}$&$4.5\times 10^{-8}$\\ 
        740 & 0.50 &0.49429  &609 &19.2  &11.7  & $2.86\times 10^{-24}$&$4.0
\times 10^{-8}$\\ 
        489 & 0.75 &0.4307   &498 & 15.7 & 12.0 & $3.77\times 10^{-24}$& $4.1
\times 10^{-8}$\\ 
230 & 0.90 &0.42043  &455 & 14.4 & 12.1 & $5.57\times 10^{-24}$ & 
$7.1\times 10^{-8}$\\   
\end{tabular}
\end{center}
\end{table}

\begin{table}[ht]
\caption{The same of Table \ref{table2} in the case of two 31 ton Mo ($v=5700$ m/s) 
hollow spheres.}
\label{table3}
\begin{center}
\begin{tabular}{cccccccc} 
\vspace*{0.1cm}
 $\Phi$ (m)& $\zeta$ & $H_1$ & $f_{0}$ (Hz) & $\Delta_{f_{0}}$ (Hz) 
& $\Gamma_{\xi}(f_{0})$ & $\sqrt{S_n(\xi;\,f_{0})}({\rm Hz}^{-1/2})$  
&$h_{0}^2\Omega_{\gamma_{\rm Edd}}$ \\[5pt]
\hline \rule{0ex}{3.0ex}
     1.82& 0.25 &0.73727  &3027 &95.7  &0.1  & $9.5\times 10^{-24}$  & 
$2.3\times 10^{-3}$\\ 
       1.88 & 0.50 &0.49429  &2304 &72.9  &3.5  & $1.1\times 10^{-23}$& 
$6.2\times 10^{-5}$\\ 
        2.16 & 0.75 &0.4307   &1650 & 52.2 & 7.2 & $1.2\times 10^{-23}$& 
$1.5\times 10^{-5}$\\ 
       2.78 & 0.90 &0.42043  &1170 & 37.0 & 9.6 & $1.3\times 10^{-23}$& 
$4.8\times 10^{-6}$\\ 
\end{tabular}
\end{center}
\end{table}

\begin{table}[ht]
\caption{The same of Table \ref{table2} in the case of two solid spheres made of CuAl 
($v$=4700 m/s) and Al5056 ($v$=5440 m/s).} 
\label{table4}
\begin{center}
\begin{tabular}{cccccccc} 
\vspace*{0.1cm}
         & $M$ (ton)& $\Phi$ (m) & $f_{0}$ (Hz) & $\Delta_{f_{0}}$ (Hz) 
& $\Gamma_{\xi}(f_{0})$ & $\sqrt{S_n(\xi;\,f_{0})}({\rm Hz}^{-1/2})$   
&$h_{0}^2\Omega_{\gamma_{\rm Edd}}$ \\[5pt]
\hline \rule{0ex}{3.0ex}
CuAl   & 105 & 3  &1672  &52.9  &7.1 & $5.0\times 10^{-24}$&  $2.6\times 10^{-6}$\\ 
       & 167 & 3.5 &1433 &45.3  &8.3  & $4.0\times 10^{-24}$&  $9.2\times 10^{-7}$\\ 
       & 250 & 4  &1254  & 39.7 & 9.2 & $3.2\times 10^{-24}$&  $4.0\times
         10^{-7}$\\[5pt] 
\hline \rule{0ex}{3.0ex}
Al5056    &  38& 3 &1935 &61.2  &5.6  & $7.2\times 10^{-23}$& $1\times 10^{-5}$\\ 
       & 60 & 3.5  &1658 &52.4  &7.1 & $5.7\times 10^{-24}$&$3.2\times 10^{-6}$\\ 
       & 90 & 4  &1451   &45.9 & 8.3 & $4.7\times 10^{-24}$&$1.3\times 10^{-6}$ \\ 
\end{tabular}
\end{center}
\end{table}

Similar analysis of correlations can be repeated for the quadrupole vibrational modes 
of the resonant spheres, which can be excited by both spin 0 and spin 2 waves.

Let us face the calculation of the overlap reduction function. The 
quadrupole tensors ${\D}^{(\epsilon)}_{ij}$ are traceless, therefore the only 
non vanishing terms in the overlap reduction function, for any 
$(\epsilon,\epsilon')$, are 
\be
\label{quadrupolo}\Gamma^{(\epsilon\epsilon')}_w(\tau)=\pi\biggl[2B_w(\tau)
{\tr}\left({\D}^{(\epsilon)}_1\,{\D}^{(\epsilon')}_2\right)+
4D_w(\tau){\tr}\left({\cal S}\,{\D}_{1}^{(\epsilon)}\,{\D}^{(\epsilon')}_{2}
\right)+E_w(\tau){\tr}\left({\cal S}\,{\D}^{(\epsilon)}_{1}\right){\tr}\left({\cal 
S}
\,{D}^{(\epsilon')}_{2}\right)\biggr]\;,
\ee
So, the functions to be inserted in 
the SNR in order to take into account all the cross correlations between 
the two spheres are 
\be
\Gamma_w(\tau)=\sqrt{\sum_{\epsilon=\epsilon'}
\left[\Gamma^{(\epsilon\epsilon')}_w(\tau)\right]^2+
\frac{1}{2}\sum_{\epsilon\neq\epsilon'}
\left[\Gamma^{(\epsilon\epsilon')}_w(\tau)
\right]^2}.
\ee

\begin{figure}[htb]
\caption{Correlation between the quadrupole modes of two spheres, one located
at the site of AURIGA (45.35 N, 11.95 E) and the other at that of NAUTILUS 
(41.80 N, 12.67 E), $d=400$ km:
 comparison between $\Gamma_h(f)$ (solid line), 
and $\Gamma_{\xi}(f)$ (dashed line).}
\label{fig2}
\begin{center}
\includegraphics[width=200 pt, height=130 pt]{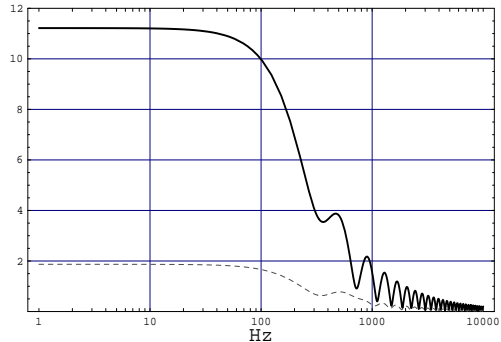}
\end{center}
\end{figure}

In Fig.\ref{fig2} we compare the plots of $\Gamma_{h}(f)$ and $\Gamma_{\xi}(f)$.
The former was computed in Ref. \cite{Coccia}.
We choose  $d=400$ km, the distance 
between the  sites of AURIGA and NAUTILUS.

\subsection{Sensitivity of VIRGO with a resonant sphere}
In this subsection we evaluate the minimum detectable scalar spectrum correlating
the monopole mode of a sphere and an interferometer, 
with the noise power spectrum of VIRGO. 
We label ${\D}_{\,1}$ the tensor of 
the monopole mode of the sphere and ${\D}_{\,2}$ the one of the 
interferometer. Since ${\tr}({\D}_1)=3/2$ and ${\D}_2$ is traceless,
the general expression for the overlap reduction function is
\be
\Gamma_{\xi}(\tau)=\frac{\pi}{2}\biggl[3C_{\xi}(\tau)+4D_{\xi}(\tau)+
E_{\xi}(\tau)\biggr]{\tr}\left({\S}\,{\D}_2\right)\;.
\ee

Explicit evaluation shows the overlap to be 
\be\label{ovsfint}
\Gamma_{\xi}(f)=4\pi j_{2}(\tau)\,{\tr}({\S}\,{\D}_{2})\;.
\ee
As pointed out in Ref. \cite{maggiore}, this function is a product of a part 
depending only on the distance $d$ and on the frequency and a part depending 
on the relative position and orientation of the detector frames. For an explicit 
estimate of the minimum detectable scalar spectrum we need the expression of 
the interferometer detector tensor with respect to the Earth centered reference 
frame, so that (\ref{ovsfint}) writes in terms of the latitude and longitude of 
the antennas. The interferometer tensor is
\be\label{interferometro}
{\D}_{\,2}^{\,ij}=\frac{1}{4}\biggl[\left(\cos 2\chi-\cos 2\psi\right)
e_{+}^{ij}(\z)+
\left(\s 2\chi-\s 2\psi\right)e_{\times}^{ij}(\z)\biggr]\;,
\ee
where $\chi$ and $\psi$ are the orientations of the two interferometer 
arms measured counterclockwise from the true North. Therefore, the location 
dependence of this tensor is split into a part depending on the position 
of the interferometer on the Earth surface and a part depending on the 
orientations of the arms with respect to the true North\footnote{Notice that 
the computations of Ref. \cite{maggiore} were performed in the interferometer 
frame and not in the Earth centered one.}.

\begin{figure}[ht]
\caption{The overlap reduction function $\Gamma_{\xi}$ of VIRGO 
(43.63 N, 10.50 E; $\chi=71.5$ deg, $\psi=341.5$ deg) 
with a resonant sphere located at: Frascati \hbox{(41.80 N, 12.67 E), $d$=270 km} 
(solid line); \hbox{(43.2 N, 10.9 E), $d$=58 km} (dashed line); Gran Sasso laboratory
\hbox{(42.4 N, 13.70 E)}, $d$=294 km (dotted line).
See the text for further explanations.}
\label{fig3}
\begin{center}
\includegraphics[width=200 pt, height=130 pt]{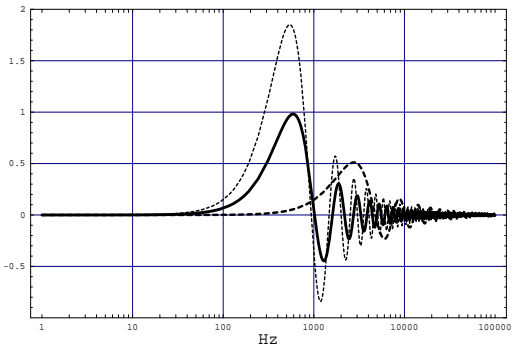}
\end{center}
\end{figure}

Since the sphere noise power spectrum is narrowbanded with respect to 
that of an interferometer, we assume the latter to be constant, and equal 
to the value of (\ref{virgospettro}) for $f = f_{n0}$, within the sphere 
bandwidth $\Delta f_{n0}$. This implies that the SNR can be written as 
\be
{\rm SNR}_{\xi}=\frac{\alpha_{0}^2}{\left(1+\alpha_{0}^2\right)}\,
\frac{3H_{0}^2}{8\pi^3}
\sqrt{\frac{\pi}{2}\Delta_{f_{n0}}T}\frac{\Omega_{\xi}(f_{n0})\Gamma_{\xi}
(f_{n0})}{f_{n0}^3\sqrt{S^{(1)}_n(f_{n0})S_{n}^{(2)}(f_{n0})}}\;.
\ee
In Fig.\ \ref{fig3} we plot $\Gamma_\xi(f)$ for $d=58$ km and $d=270$ km, 
which is the distance between VIRGO and the site of NAUTILUS in Frascati.
In this figure we plot also the curve for the case in which the sphere 
is located nearly in the Gran Sasso underground laboratory ($d=294$ km). For 
this correlation $\Gamma_\xi(f)$ gets approximately its maximum value. This 
result is particularly important in view of the remark that for the future 
resonant detectors with project sensitivity approaching the quantum limit, 
the cosmic ray interactions in the detector may set a limit to the sensitivity 
in an unshielded environment.

For $d$=270 km the function has its first peak at $\simeq 591$ Hz, with 
$\Gamma_{\xi}\sim 0.98$, and its first zero at $\simeq 1019$ Hz. Notice 
that, as the spacing $d$ increases, the peak frequency moves to lower 
values. This is peculiar of the correlation between a sphere and an 
interferometer, because for two spheres, two interferometers or two 
bars an increase of $d$ simply implies a shift to lower values of the 
first zero. For a sphere and an interferometer this effect is due 
to the function $j_{2}(\tau)$ in (\ref{ovsfint}). The relative orientation 
dependent factor, ${\tr}({\cal S}\,{\D}_{2})$, accounts for the full amplitude 
of the overlap reduction function, being an oscillating function. In fact 
we can express the overlap reduction function with respect to the natural 
frame of the interferometer defined before \cite{maggiore}. In this frame, 
the direction of the unit vector $\hat{s}$ joining the antennas is 
determined by the angles $(\theta,\phi)$ and the detector tensor has the 
simple form (\ref{Michelagnolo}). 
Using the same convention for the angles as in \cite{maggiore} we have  
\be\label{BianConiglio}
{\tr}({\S}\,{\D}_{2})=\frac{1}{2}\sin^2\theta\cos 2\phi.
\ee

In Table \ref{table5} we consider Mo and CuAl hollow spheres at $d=270$ km. 
The CuAl sphere gives the best sensitivities, because the resonance 
frequency is lower and it belongs to the range where the overlap 
reduction function gets its maximum. $S^{(1)}_n$ is the noise power 
spectrum of the sphere and $S^{(2)}_n$ that of VIRGO, both for the 
frequency $f_{10} \equiv f_0$.

\begin{table}[ht]
\caption{Minimum detectable scalar spectrum (SNR$_{\xi}=1, T=1$ year) 
obtained by the first monopole vibrational mode of one hollow sphere 
at Frascati with  VIRGO ($d=270$ km). The CuAl and Mo hollow spheres 
are those of Table \ref{table2} and Table \ref{table3}, therefore only 
their resonance frequencies and the corresponding sensitivities are written 
here.}
\begin{center}
\begin{tabular}{c  c c c c c c c } 
\vspace*{0.1cm}
       & $f_0$ (Hz) & $\Delta_{f_0}$ (Hz)
& $\Gamma_{_\xi}(f_0)$ & $\sqrt{S_n^{(1)}}({\rm Hz}^{ 1/2})$  &
$\sqrt{S_n^{(2)}}({\rm Hz}^{ 1/2})$&
$h_{0}^2\Omega_{\gamma_{\text{Edd}}}$\\[5pt]
\hline \rule{0ex}{3.0ex}
CuAl &770 &24.3  &0.76  & $2.2\times 10^{-24}$ &$4.1\times 10^{-23}$&
$1.2\times 10^{-5}$\\
    &609 &19.2  &0.98  & $2.9\times 10^{-24}$ & $3.9\times 10^{-23}$&
$6.6\times 10^{-6}$\\
    &498 & 15.7 & 0.92& $3.8\times 10^{-24}$ & $3.9\times 10^{-23}$&
 $5.6\times 10^{-6}$\\
    &455 & 14.4 & 0.85 &$5.6\times 10^{-24}$ &  $4.0\times 10^{-23}$&
$7.2\times 10^{-6}$\\[5pt]
\hline \rule{0ex}{3.0ex}
Mo  &3027 &95.7  &0.19 & $9.4\times 10^{-24}$ & $1.1\times 10^{-22}$&
$1.8\times 10^{-2}$\\
     &2304 &72.9  &0.16& $1.2\times 10^{-23}$ & $8.6\times 10^{-23}$ &
$1.0\times 10^{-2}$\\
     &1650 & 52.2 & 0.08 & $1.2\times 10^{-23}$& $6.4\times 10^{-23}$ &
$7.0\times 10^{-3}$\\
     &1170 & 37.0 & 0.35& $1.3\times 10^{-23}$&  $5.0\times 10^{-23}$&
$5.2\times 10^{-4}$\\
\end{tabular}
\end{center}
\label{table5}
\end{table}

\section{Conclusions}
The aim of this paper was to study the detectability of scalar GWs from
the cosmic stochastic \text{GW} background. Before discussing our results, 
we would briefly like to recall what is the best strategy to perform such a 
measurement in our opinion. As we argued in the introduction, 
as far as the type of detector to use is concerned, it does not seem
practical to use only  L shaped interferometers. If the
impinging monopole mode of a
scalar GW moves along the $\hat{z}$ axis and the arms of the interferometer
lie in the $\x, \y$ plane, then the two arms will be stretched by
the same quantity $\delta L$. In this case an interferometer set to work on a 
dark fringe will not detect any signal. On the contrary, in this
very configuration a spin 2 GW will give its maximum effect.
This is a limiting case though.
In general the direction of the
impinging GW will form a certain angle with $\hat z$ and the
perturbation due to a scalar will be tangled with that of the spin 2 in
an inextricable way. 
In principle one could reconstruct the directional
sensitivity patterns for the two spins to separate the two signals. The data
needed to do this will require much time to be gathered and for this reason 
also this proposal does not seem practical to us.
What would happen if the detector were a sphere?
Let us first analyze the case of a GW from the viewpoint of a reference
frame centered in the origin of the sphere.
As it was discussed in Ref. \cite{Colacino} the GW interacts with the
resonant mass detector through the so called electric tensor $E_{ij}=R_{i0j0}$.
If the direction of propagation of the incoming GW is along the $\hat{z}$ axis,
the six components of this tensor can be expressed, using a null tetrad, 
in terms of the so called Newman-Penrose parameters which
can also be expressed in the basis given by the ${\S}_{ij}^{(lm)}$
defined in Appendix \ref{appendixC1}. In turn
this basis can be put in relation with the actual measurements performed by 
the detectors \cite{Colacino}. 
In this reference frame there is a direct correspondence between
the Newman-Penrose parameters and the spin of the incoming GW and the 
information about the spin of the GW can be easily extracted. If the direction
of propagation of the GW makes an angle with the $\hat{z}$ axis, the situation is 
more complicated as it can also be seen from (\ref{tscatt1})
which is the cross section given by a GW with polarization $e^+, e^{\times}$
and from (\ref{sscatt}), which is the analogous with polarization $e^{s}$.
In (\ref{tscatt}), for example, even if the incoming wave is pure spin 2
all the modes are excited and the situation is indistinguishable from that
described by (\ref{sscatt}). The situation resembles very much that for an 
interferometer.
But in the case of the sphere, the
pure monopole  mode given from (\ref{sezione00}) is only excited by the 
scalar wave, giving a clear signal of the presence of a scalar wave.
This is the motivation for our proposal to couple an 
interferometer to a resonant detector of spherical shape. Let us now look at 
the results we have obtained that, for the sake of generality, encompass
also the sensitivities for pairs of resonant mass detectors.
 
The behavior of the overlap reduction function for pairs of resonant mass 
detectors (Fig.\ \ref{fig1}-\ref{fig2}) is quite different from the one for a resonant mass 
detector 
and an interferometer (Fig.\ \ref{fig3}). In Fig.\ \ref{fig1}-\ref{fig2} we see a constant function which
abruptly goes to zero for certain values of the frequency. 
These values of the frequency decrease by increasing the distance $d$ at 
which the two detectors are located. The values of the monopole overlap reduction
function and of $\Gamma_h(f)$ are of the same order of magnitude: however, the quadrupole 
$\Gamma_{\xi}(f)$ is
even an order of magnitude smaller at low frequencies.

Quite on the contrary, the overlap reduction function for the pair 
interferometer-resonant mass detector is different from zero only in a 
certain region which depends from the distance between the antennas and
the direction $\hat{s}$ of the sphere with respect to the arms of the interferometers.
In Fig.\ \ref{fig3} it is shown that the values of the frequencies at which the overlap
is maximum are in agreement with the resonant frequencies of the planned 
detectors \cite{odia}.

The numerical results concerning the sensitivities are given in 
Tables \ref{table2}, \ref{table3}, \ref{table4} for pairs of resonant mass detectors 
and in Table \ref{table5} for the pair interferometer-resonant mass detector.

The values given in Tables \ref{table2}, \ref{table3} show the potential of hollow spheres:
going from realistic weights for such detectors of the order of the dozens of tons to
weights of the order of the hundreds of tons (which are non realistic at the 
present state of the art) there is a gain in sensitivity of two orders of magnitude.
Such a gain could also be achieved going to materials with a higher speed of 
sound propagation \cite{Cerdo} as can be seen from (\ref{sensitivity0s}) where 
both $M$ and $v$ appear in the denominator but $v$ is squared.

How do these results compare to those obtained for spin 2 \text{GW}? It depends on the value
of the scalar amplitude $\alpha_0^2$ and the scalar coupling $\tilde q^a$
since $\Omega_\xi$ is roughly proportional to the 
scalar amplitude and coupling while  $\Omega_h$ is roughly proportional 
to $1+\alpha_0^2$ (see (\ref{SNR}) and (\ref{SNRh})). At the moment $\alpha_0^2$ has been
only measured in our solar system and its value at 1$\sigma$ level is 
$\alpha_0^2\approx 10^{-3}$ \cite{Lebach}.
Such a small value is given to the fact that Einsteinian general relativity 
seems to be very well verified. Such a value of $\alpha_0^2$ would give very little chances to the planned
resonant detectors to detect a scalar \text{GW} background which should be limited by
nucleosynthesis to be \cite{Schwartz}
\be\label{Nucleosintesi}
\int_{f>10^{-8}Hz}h_{0}^2\Omega_{\xi}(f){\d}(\ln f)<10^{-5}.
\ee
We have also to mention Ref. \cite{santiago}, where an estimate of $\alpha_0^2$ was attempted
starting from the same nucleosynthesis bound: the result is a weak dependence of 
$\alpha_0^2$ from distance. Our ignorance on the mechanisms that should give $\alpha_0^2$
its value (cosmological attractor? supersymmetry breaking?) prevents us from further comments
on this point. Also concocting strategies with resonant mass detectors made of different
materials (to exploit the $\tilde q_a$ dependence) is possible, but probably premature
given what we said earlier. We remark however that once operating, resonant mass detectors 
of spherical shape could themselves provide a measure of $\alpha_0^2$ using binary 
or collapsing stars as emphasized by many authors and more in particular 
in Ref. \cite{Cocciafucito}.

As a final comment we remark that the sensitivity of the pair interferometer-resonant mass 
detector seems to be a couple of orders of magnitude less than that of a pair of resonant 
mass detectors. The plots we have given show that a careful choice of where to locate the 
detectors can account for up to an order of magnitude in sensitivity.

\section*{Acknowledgments}
We would like to thank M. Gasperini for useful discussions. A.N. would like 
to thank Andrea Nagar 
for very helpful suggestions and Bobet Klein for many encouragements during the 
completion of this work.

\newpage
\appendix
\section{Correlation functions for many scalar fields}\label{appendixA}
The SNR for scalar GWs has been 
computed in Ref. \cite{maggiore} following Ref. \cite{Allen}. Actually, 
although the generalization to single scalar theories (Brans-Dicke) is 
trivial, 
the one to multi scalar theory needs the introduction of a very strong 
constraint 
on the fields themselves: in order to get a formula for the SNR
one can state the following lowest order condition for the 
correlation function between the Fourier amplitudes of the scalar fields 
\be
\label{correlatore}<\xi^{a*}(f,\O)\xi^{b}(f',\O)>=\gamma^{ab}_0\,
\delta(f-f')\delta(\O-\O')
K(f)\;,
\ee
where $K(f)$ is a real non negative symmetric function. 
This hypothesis 
means 
that the correlation function is the same for every pair
of scalar fields.
This is not the most general situation one can imagine:
in fact, because of the symmetry $a\leftrightarrow b$, one would expect 
to have $n(n+1)/2$ distinct correlation functions $K_{ab}(f)$. On the 
other hand, 
if we consider only one degenerate function $K(f)$, then the framework is exactly 
the same of Refs. \cite{maggiore,Allen}, and, with the same algebra, we get 
(\ref{SNR}).

We reproduce the main steps to express $K(f)$ in terms of the spectrum $\Omega_{s}$. 
Straightforward generalization of \cite{maggiore} shows that, for any tensor 
multi scalar theory, the energy density carried by a GW is 
\be
\tau_{00}=\rho_{h}+\rho_{s}=\bigl(1+\alpha_0^2\bigr)\left(32\pi 
\tilde{G}\right)^{-1}\biggl[
<\dot{h}_{\mu\nu}\dot{h}^{\mu\nu}>+\,8\gamma^{0}_{ab}
<\dot{\xi}^a\dot{\xi}^b>\biggr],
\ee
where the brackets $<\dots>$ stand for integration over a finite region of 
tridimensional space containing several wavelengths. From this formula we recover
 $K(f)$ as a function of the scalar spectrum $\Omega_s(f)$. From (\ref{correlatore}) 
 one obtains 
\be
<\dot{\xi}^{a}\dot{\xi}^{b}>=32\,\pi^3\,\gamma^{ab}_{0}\int_{0}^{\infty}df\,f^2\,K(f).
\ee 
Using the definition of $\Omega_{s}$, for non negative $f$, we get
\be
\Omega_{s}(f)=\frac{f}{\rho_c}\frac{{\d}\rho_s}{{\d}f}=n\,(1+\alpha_0^2)\,\frac{64\pi^3}
{3 H_{0}^2}\,f^3\,K(f),
\label{app1}
\ee
where we used $\gamma_{ab}^0\,\gamma^{ab}_0=n$. 
Furthermore $\Omega_{s}=\sum_{a}\Omega_{\xi^a}
=n\Omega_{\xi^a}$ so we can infer from (\ref{app1})
\be
K(f)=\left(1+\alpha_{0}^2\right)^{-1}\frac{3H_{0}^2}{64\pi^3}\,f^{-3}\,\Omega_{\xi}(f),
\ee
where $\Omega_{\xi^{a}}\equiv\Omega_{\xi}$.

\section{The overlap reduction functions}\label{appendixB}
In the following we give the coefficients, introduced in the main text,
for the functions $\Gamma_{\xi}(f)$
\be
\pmatrix{
A\cr B\cr C\cr D\cr E}_{\xi}
=\frac{4}{\tau^2}\
\pmatrix{\tau^2\,j_{0}(\tau)-
2\,\tau j_{1}(\tau)+j_{2}(\tau)\cr
j_{2}(\tau)\cr -\tau^2\,j_{0}(\tau)+
4\,\tau\, j_{1}(\tau)-5\,j_{2}(\tau)\cr
\tau\,j_{1}(\tau)-5\,j_{2}(\tau)\cr
\tau^2\,j_{0}(\tau)-
10\,\tau\,j_{1}(\tau)+35\,j_{2}(\tau)}\;,
\ee
and $\Gamma_{h}(f)$ 
\be
\pmatrix{A\cr B\cr C\cr D \cr E}_{h}=
\frac{4}{\tau^2}
\pmatrix{-\tau^2 j_{0}(\tau)-2\,\tau j_{1}(\tau)+j_{2}(\tau)\\
\tau^2 j_0(\tau)-2\tau j_1(\tau)+j_{2}(\tau)\cr
-\tau^2 j_{0}(\tau)-2 \tau j_{1}(\tau)-5 j_{2}(\tau)\cr
-\tau^2 j_0(\tau)+4\tau j_{1}(\tau)-5 j_{2}(\tau)\cr
\tau^2 j_{0}(\tau)-
10 \tau j_{1}(\tau)+35 j_{2}(\tau)}\;.
\ee

\section{Detector tensors}\label{appendixC}

\subsection{The sphere mode tensors}\label{appendixC1}
A basis for the pure spherical harmonics is given by the ${\S}^{(lm)}$, 
with $l=0,2$ \cite{Lobo} 
\bea
{\S}^{(00)}&\equiv&\frac{1}{\sqrt{4\pi}}
\pmatrix{
1 & 0 & 0\cr
0 & 1 & 0\cr
  0 & 0 & 1},\nonumber\\
{\S}^{(20)}&\equiv&\sqrt{\frac{5}{16\pi}}
\pmatrix{-1 & 0 & 0\cr
0 & -1 & 0\cr
0 & 0 & 2},\nonumber\\
{\S}^{(2\pm2)}&\equiv&\sqrt{\frac{15}{32\pi}}\pmatrix{
1 & \pm i & 0\cr
\pm i & -1 & 0\cr
0 & 0 & 0},\nonumber\\
{\S}^{(2\pm1)}&\equiv&\sqrt{\frac{15}{32\pi}}\,
\pmatrix{
0 & 0 & \mp 1\cr
0 & 0 & -i\cr
\mp 1 & -i & 0}.
\eea
The normalization is chosen so that 
${\S}_{ij}^{(lm)}\hat{n}^{i}\hat{n}^j=Y_{lm}$. 
$\hat{n}$ is the radial unit vector. 
The vibrational response 
of a spherical detector is usually written in terms of this pure spin basis.
Otherwise, following Zhou and Michelson \cite{Zhou}, the vibrations of a resonant sphere 
are more conveniently described as functions of the real quadrupole spherical 
harmonics, in addition to the monopole spherical harmonic
$Y_{00}=\left(4\pi\right)^{-1/2}$
\bea
\nonumber{Y_{0}}&\equiv &Y_{20}\; ,\\
\nonumber{Y_{1c}}&\equiv&\frac{1}{\sqrt{2}}\left(Y_{2-1}-Y_{2+1}\right)\;,\\
Y_{1s}&\equiv&\frac{i}{\sqrt{2}}\left(Y_{2-1}+Y_{2+1}\right)\;,\\
\nonumber{Y_{2c}}&\equiv&\frac{1}{\sqrt{2}}\left(Y_{2-2}+Y_{2+2}\right)\;,\\
\nonumber{Y_{2s}}&\equiv&\frac{i}{\sqrt{2}}\left(Y_{2-2}-Y_{2+2}\right)\;.
\eea
A convenient basis for the real spherical harmonics is given by 
${\D}^{(00)}\equiv \sqrt{\pi}{\S}^{(00)}$ and ${\D}^{(\epsilon)}$ with 
$\epsilon\equiv 0,1c,1s,2c,2s$. These traceless tensors 
are defined as  
\bea \label{sphere real tensors}
{\D}^{(0)}&\equiv&\frac{\sqrt{3}}{6}
\pmatrix{
1 & 0 & 0\cr
0 & 1 & 0\cr
0 & 0 & -2},\nonumber\\
{\D}^{(1c)}&\equiv&-\frac{1}{2}\pmatrix{
0 & 0 & 1\cr
0 & 0 & 0\cr
1 & 0 & 0},\nonumber\\ 
{\D}^{(1s)}&\equiv& -{1\over 2}
\pmatrix{
0 & 0 & 0\cr
0 & 0 & 1\cr
0 & 1 & 0},\\ 
{\D}^{(2c)}&\equiv&
{1\over 2}\pmatrix{1 & 0 & 0\cr
0 & -1 & 0\cr
0 & 0 & 0},\nonumber\\
{\D}^{(2s)}&\equiv&
-{1\over 2}\pmatrix{
0 & 1 & 0\cr
1 & 0 & 0\cr
0 & 0 & 0 },\nonumber
\eea
with
\be
\sum_{i}\sum_{j}{\D}_{ij}^{(\epsilon)}{\D}_{ij}^{(\epsilon')}=\frac{1}{2}\,
\delta^{\epsilon\epsilon'}\;.
\ee
From the definition of the real spherical harmonics we have
\bea\label{tensorsfera}
\nonumber{{\D}^{(0)}}&=&-\sqrt{\frac{4\pi}{15}}\,{\S}^{(20)}\;,\\
\nonumber{{\D}^{(1c)}}&=&\sqrt{\frac{2\pi}{15}}\,\left({\S}^{(2+1)}-
{\S}^{(2-1)}\right)\;,\\
{\D}^{(1s)}&=&-i\,\sqrt{\frac{2\pi}{15}}\,\left({\S}^{(2+1)}+{\S}^{(2-1)}
\right)\;,\\
\nonumber{{\D}^{(2c)}}&=&\sqrt{\frac{2\pi}{15}}\,\left({\S}^{(2+2)}+
{\S}^{(2-2)}\right)\;,\\
\nonumber{{\D}^{(2s)}}&=&i\,\sqrt{\frac{2\pi}{15}}\,\left({\S}^{(2+2)}-
{\S}^{(2-2)}\right)\;.
\eea

\subsection{Explicit expressions of the polarization
tensors in the detector frame}\label{appendixC2}

Let us consider now the wave frame $(\m,\n,\O)$ and the detector frame $(\x,\y,\hat{z})$ 
defined 
in (\ref{rotaziuncella}) by introducing the rotation matrix 
\be\label{matrot}
R(\O)\equiv
 \pmatrix{
\cos\phi & \sin\phi & 0\cr
-\cos\theta\sin\phi & \cos\theta\cos\phi & \sin\theta\cr
\sin\theta\sin\phi  & -\sin\theta\cos\phi & \cos\theta}, 
\ee
where the angles $\left(\theta,\phi\right)$ are defined following the conventions of 
Forward \cite{Forward}. The polarization tensors of the GW in the antenna 
frame $e^{B}(\O)$ are obtained by rotating the ones in the wave frame $e^B$ as
\be\label{transf}
e^{B}(\O)=R^{t}(\O)\,e^{B}\,R(\O);\qquad\quad B=\times,+,{s}.
\ee
The tensors in the wave frame are 
\be
{e}^{+}=
 \pmatrix{
                     1 & 0 & 0\cr
                     0 & -1 & 0\cr
                     0 & 0 & 0
                     }, 
\qquad
{e}^{\times}=
 \pmatrix{
                     0 & 1 & 0\cr
                     1 & 0 & 0\cr
                     0 & 0 & 0
                     },
\qquad
 {e}^s =
 \pmatrix{
                     1 & 0 & 0\cr
                     0 & 1 & 0\cr
                     0 & 0 & 0
                     }.                    
\ee
From (\ref{transf}),
for the spin 2 polarization tensors we get 
\be
{e}^{+} (\O)=\m\otimes\m-\n\otimes\n=\frac{1}{2}
 \pmatrix{
                     2(\cos^2\phi-cos^2\theta\sin^2\phi) & (1+\cos^2\theta)
\sin2\phi                  & \sin 2\theta\sin\phi\cr
                     (1+\cos^2\theta)\sin2\phi     & 2(\sin^2\phi-\cos^2\theta\cos^2\phi) & -\sin2\theta\cos\phi\cr
                     \sin 2\theta\sin\phi          & -
\sin2\theta\cos\phi                       & -2\sin^2\theta
                     }, 
\ee
\be
{e}^{\times}(\O)=\m\otimes\n+\n\otimes\n= \pmatrix{
                        -\cos\theta\sin 2\phi        & \cos\theta\cos 2\phi   & 
\cos\phi\sin\theta\cr
                        \cos\theta\cos 2\phi         & \cos\theta\sin 2\phi   & 
\sin\theta\sin\phi\cr
                        \cos\phi\sin\theta           & \sin\theta\sin\phi     & 0 
 }. 
 \ee
 For the scalar polarization tensor we have 
\be
{e}^{s}(\O)=\m\otimes\m+\n\otimes\n= \frac{1}{2}
 \pmatrix{
       2(\cos^2\phi+\cos^2\theta\sin^2\phi)   &  \sin^2\theta\sin 2\phi                   
       & -\sin 2\theta\sin\phi\cr
        \sin^2\theta\sin 2\phi       &  2(\sin^2\phi+\cos^2\theta\cos^2\phi) &  
\sin 2\theta\cos\phi\cr
        -\sin 2\theta\sin\phi        &  \sin 2\theta\cos\phi                     & 
2\sin^2\theta}. 
\ee

\subsection{The Earth centered reference frame}\label{appendixC3}

In Appendix \ref{appendixB} we have listed the coefficients which give the dependence
of the overlap reduction function on the frequency and on the distance between 
the antennas. To infer its dependence on the relative orientations of 
the detectors, it is convenient to express the detector tensors in the reference 
frame of the Earth. We then express the detector tensors with respect to 
a triad of orthogonal unit vectors $(\x,\y,\z)$, where $\x$ and $\y$ 
lie on the tangent plane and $\z$ points along the Earth radius. This 
triad defines univocally the antenna coordinate system. 
Given the latitude, $\Theta$, measured in degrees North from the equator 
and the longitude, $\Phi$, in degrees East of Greenwich, England, 
the relation of the triad of vectors $(\x,\y,\z)$ with respect to the 
Cartesian reference 
frame $(\hat X,\hat Y,\hat Z)$ originated in the center
of the  Earth is
\bea
\hat x&=&\;-\sin\Theta\cos\Phi\X-\sin\Theta\sin\Phi\Y
+\cos\Theta\Z ,\nonumber\\
\y&=&-\sin\Phi\X+\cos\Phi\Y ,\\
\z&=&\cos\Theta\cos\Phi\X+\cos\Theta\sin\Phi\Y+\sin\Theta\Z\nonumber\;.
\eea

A simple example is given by the tensor of an interferometer
 \cite{Forward} which, in the Earth centered frame, is usually written as
\be
{\D}_{ij}(\X,\Y)=\frac{1}{2}(\X_i\X_j-\Y_i\Y_j)\;,
\ee
where $\X$ and $\Y$ are chosen to point in the detector arms 
directions.

Frame dependent expressions of the same kind can also be written 
for the tensors describing the geometrical features of the modes 
of a resonant sphere. In the Earth centered reference frame, 
equations (\ref{tensorsfera}) become
\bea
{\D}^{(0)}_{ij}&=&\frac{\sqrt{3}}{6}\left(\X_i\X_j+\Y_i\Y_j-2\Z_i\Z_j\right),
\nonumber\\
{\D}^{(1c)}_{ij}&=&-\frac{1}{2}\left(\X_i\Z_j+\Z_i\X_j\right),
\nonumber\\
{\D}^{(1s)}_{ij}&=&-\frac{1}{2}\left(\Y_i\Z_j+\Z_i\Y_j\right),\\
{\D}^{(2c)}_{ij}&=&\frac{1}{2}\left(\X_i\X_j-\Y_i\Y_j\right),
\nonumber\\
{\D}^{(2s)}_{ij}&=&-\frac{1}{2}\left(\X_i\Y_j+\Y_i\X_j\right).\nonumber
\eea

\end{document}